\documentclass[twocolumn, showpacs]{revtex4}
\usepackage{CJK}
\usepackage{graphics}
\usepackage{amsmath}
\usepackage{mathrsfs}
\usepackage{amsfonts}
\usepackage{overpic}
\usepackage{hyperref}
\usepackage{rotating}
\usepackage{amssymb}
\usepackage{color}
\newcommand{\ket}[1]{|#1\rangle}
\newcommand{\bra}[1]{\langle #1|}
\newcommand{\inp}[2]{\langle #1 | #2\rangle}

\begin{document}
\begin{CJK*}{GB}{gbsn}
\title{ Maximal Overlap with the Fully Separable State
and Translational Invariance for Multipartite Entangled States}
\author{H. T. Cui(´Þº£ÌÎ)\footnote{Electric address: cuiht@aynu.edu.cn}, Di Yuan(Ô¬µØ) and J. L. Tian(Ìï¿¡Áú)}
\affiliation{School of Physics and Electrical Engineering, Anyang
Normal University, Anyang 455000, China}
\date{\today}
\begin{abstract}
The maximal overlap with the fully separable state for the
multipartite entangled pure state with translational invariance is
studied explicitly by some exact and numerical evaluations, focusing
on the one-dimensional qubit system and some representative types of
translational invariance. The results show that the translational
invariance of the multipartite state could have an intrinsic effect on
the determinations of the maximal overlap and the nearest fully
separable state for multipartite entangled states. Furthermore a
hierarchy of the basic entangled states with translational
invariance is founded, from which one could readily find the maximal
overlap and a related fully separable state for the multipartite state
composed of different translational invariance structures.
\end{abstract}
\pacs{03.65.Ud} \maketitle
\end{CJK*}

\section{introduction}
Quantum entanglement is considered as the most distinct feature in the
quantum world to the classical world. Generally it is the
manifestation of the \emph{nonlocal} connectedness in the
distinguishable parties, where the nonlocal means that if only this
connectedness is constructed, it does not disappear automatically
until the appearance of decoherence or local measurements, even if
the parties are space-like separated from each other. An important
consequence of this connectedness is that the behaviors of any one
party are inevitably affected by the other parties. Thus it is
convenient for the quantification of the connectedness to find the
reduced density matrix of the subsystem. Some successful criteria
or measures of quantum entanglement have been proposed along this
line (see Ref \cite{Horodecki} for a comprehensive review).

It should be pointed out that the intuition of quantum entanglement
is mainly from the understanding of Bell states, which are maximally
entangled states of two-qubit systems. Particularly because Bell
states are bipartite, one can obtain complete information on the
entanglement from the reduced density matrix of the subsystem.
However the situation becomes complex when extended to multipartite
systems. Multipartite entanglement could exist when the multipartite
state \emph{cannot} be written as the fully separable form
\begin{equation}\label{msep}
\rho^{sep}_f=\sum_i p_i
\rho^{(i)}_1\otimes\rho^{(i)}_2\otimes\cdots\rho^{(i)}_N,
\end{equation}
where $N$ is the number of the distinguishable party, and $p_i$
denotes the joint probability distribution of the single-party state
$\rho_n^{(i)}(n=1,2,\cdots,N)$. In contrast to the bipartite
entanglement defined as the violation of the bi-separable form for
the bipartite state $\rho^{sep}_b=\sum_i p_i
\rho^{(i)}_1\otimes\rho^{(i)}_2$, there exists the
multi-connectedness in multipartite states, which is hard to
characterize completely by the reduced density matrix of the
subsystem only.

This distinction between bipartite and multipartite entanglement
has led to several intrinsic observations. It is known that there
exist two inequivalent three-qubit entangled states, the $W$ state and
Greenberger-Horn-Zeilinger (GHZ) state, which are not
interconvertible under local operations and classical communications
(LOCCs) \cite{dvc00}. Furthermore, a limit to the distribution of
entanglement in multipartite states is found first for the three-qubit
case \cite{ckw00}, and then for the arbitrary multipartite case
\cite{ov06}. These phenomena imply that one need some special
methods to describe the connectedness embedded in multipartite
states.

The global approach is a natural choice to obtain the comprehensive
information of the connectedness in multipartite states. In contrast
to the dependence of the measurements of bipartite entanglement on the
subsystem, the global measurement of multipartite entanglement instead focuses
mainly on the overall state instead. For instance the relative
entropy of entanglement is one type of measurement, defined as
\cite{vb98}
\begin{equation*}
E_R=\min_{\{\rho^{sep}\}}\text{Tr}[\rho(\log\rho-\log\rho^{sep})].
\end{equation*}
The key idea in this measurement is that the closer to the separable
states is $\rho^{sep}$, the less entangled is $\rho$. Global robustness
is another global measure of entanglement, of which the main idea is
to quantify how robust is the entangled state against the
environmental noise \cite{hn03}. In addition there exists a global
criterion of entanglement: entanglement witness. The idea is to find
a special Hermitian operator, whose expectation value with the
multipartite state is positive or zero when this state is fully
separable, while it is negative when this state is entangled \cite{hhh}.

Similar to the relative entropy of entanglement, geometric
entanglement (GE) is another global measure of entanglement, related
directly to the distance between the entangled state and the fully
separable state in Hilbert space. GE is defined generally for the pure
state as \cite{wg03}
\begin{equation}
E_g=\min_{\{\ket{\phi}\}}\|\ket{\psi}-\ket{\phi}\|^2,
\end{equation}
where $\|\cdots\|$ denotes the norm, or equivalently
\begin{equation}
E_g=1-\Lambda^2_{\max}=1-\max_{\{\ket{\phi}\}}\left|\inp{\psi}{\phi}\right|^2.
\end{equation}
where $\ket{\phi}=\otimes_{i=1}^N\ket{\phi^i}$ is a fully separable
pure state,  $\ket{\phi^i}$ denotes the single-party state, and
$\Lambda_{\max}$ denotes the maximal overlap of $\ket{\psi}$ and
$\ket{\phi}$.  $E_g$ is determined geometrically by the overlap
angle between the state vectors $\ket{\psi}$ and $\ket{\phi}$ in
Hilbert space. Thus the optimal in the definitions above can be
reduced to find the nearest fully separable state $\ket{\phi}$,
which has a minimal overlap angle with an entangled state
$\ket{\psi}$. Furthermore, the determination of the nearest
$\ket{\phi}$ is equivalent physically to the determination of the
Hartree-Fock approximation ground state of the auxiliary Hamiltonian
$H=-\ket{\psi}\bra{\psi}$
 \cite{wg03}. Additionally our recent study shows that
$\Lambda_{\max}$ would be a direct generalization of the concept of
Anderson orthogonality catastrophe (AOC) in solid theory
\cite{cui10}, and thus GE can be used to also describe the
correlations in many-body systems \cite{br07, wdmvg05, cwy10}. This
point can be manifested by the expression
\begin{equation}\label{aoc}
\Delta=|\inp{\Phi}{\Phi^p}|^2,
\end{equation}
where $\ket{\Phi}$ corresponds to the true ground state of a many-body
system, and $\ket{\Phi^p}$ is actually a pure product state
described entirely in terms of free plane waves, which can be
considered as the ground state without potential \cite{anderson67}. AOC
denotes the vanishing of $\Delta$ under a thermodynamic limit, even
for a very weak potential. This feature discloses that the correlation
in many-body systems is the intrinsic character that is different from the
free system without potential. Then, by finding the tendency of
$\Delta$ under thermodynamic limit, one can obtain  information
on the correlation in systems. With these points, GE actually defines a
measurement of the correlation in multipartite states, independent of
the details of the system because of the optimal choice of the fully
separable state. Furthermore because of the optimal the measured
correlation is evidently quantum, as proved in the next section.

These distinct characters display the popularity of GE as an
description of the connectedness in multipartite states. Also GE has
become one of the most accepted measures of multipartite
entanglement. However, it is difficult in general to find the maximal
overlap $\Lambda_{\max}$ because of the optimal of $\ket{\phi}$,
which is the crucial point for the evaluation of GE, and there are
few exact results \cite{wg03,wdmvg05, cwy10,cw07, odv08, ws09,
hkwgg09,hmmov09, tst10, mgbbb10, ow10, sapfv10}. Recently
important progress has been made for the entangled state with
permutational invariante that the nearest fully separable state for
it is \emph{necessarily} permutational invariance \cite{ws09,
hkwgg09,hmmov09}; that is to say, it is always an optimal choice
to set $\ket{\phi}=\ket{\phi'}^{\otimes N}$ in order to find
$\Lambda_{\max}$ of this type of entangled state, in which
$\ket{\phi'}$ denotes the single-party state. This important conclusion
implies strongly that it would reduce the optimal determination of
GE by utilizing the symmetry of $\ket{\psi}$. Furthermore our recent
study also shows that the maximal overlap could be obtained if
the fully separable state, either pure or mixed, shows the same global
symmetry to the entangled state \cite{cui10}, which means,
mathematically, that the two states should belong to the same
symmetric subspace.

However, except for permutationally invariant entangled states,
there are few examples for the evaluations of GE for other
multipartite states. This article serves to fill this gap partially.
For this purpose, GE for the multipartite states with translational
invariance is studied explicitly through some exact and numerical
examples, which are focused on the qubit system with the geometry of circles and
some representative types of translational invariant entangled
states. Another reason for this choice comes from revisiting the
current points of the determination of $\Lambda_{\max}$ for
translationally invariant states, appearing in some very recent
works; for examples, one case is that the nearest fully separable
state could not be determined  by the translational invariance of
the entangled state, and thus this symmetry would be helpless for the
reduction of the optimal in GE \cite{hkwgg09}. One case is to set
$\ket{\phi}=\ket{\phi'}^{\otimes N}$ in order to obtain
$\Lambda_{\max}$ of the ground state in translationally invariant
many-body systems \cite{wg03,ow10}. Another case is to adopt the
maximal coefficient under the product state basis as the maximal
overlap \cite{odv08}, etc.

Through several exact and numerical calculations, we try to
illustrate convincingly in this article that the optimal
determination of the nearest fully separable state and maximal
overlap with a fully separable state for translationally invariant
entangled states can be reduced greatly by utilizing the
translational invariance of the entangled state. Additionally, the above
points can occurs only for some special cases. Furthermore our
study shows that there exists a hierarchy for the so-called basic
translationally invariant entangled states defined in Sec.\ref{sec:
overlapII}, from which one can decide directly the nearest fully
separable state and the $\Lambda_{\max}$.

\section{technical preparations}
Some concepts are clarified in this section. At the end of
this section we present a proof for the point that for an entangled
pure state there always exists a nearest fully separable pure state.

\emph{Permutational invariance} (PI) denotes formally the situation
that the multipartite state is unchanged by exchanging the states of
two arbitrary single parties. For example, the $N$-qubit GHZ state,
\begin{equation}
\ket{\text{GHZ}}_N=\tfrac{1}{\sqrt{2}}(\ket{11\cdots1}+\ket{00\cdots0})
\end{equation}
is obviously permutationally invariant because all parties always have
the same state simultaneously. In contrast, the generalized $W$ state,
\begin{equation}
\ket{W}_N=\tfrac{1}{\sqrt{N}}(\ket{10\cdots0}+\ket{010\cdots0}+\cdots+\ket{0\cdots01})
\end{equation}
is slightly special; although it is also permutationally invariant,
the key feature is that $\ket{W}_N$ includes all possible
combinations of the single $\ket{1}$ and the $(N-1)$'s $\ket{0}$.
This difference would induce a distinct hierarchy from
$\ket{\text{GHZ}}_N$, as shown in Sec.\ref{sec: overlapII}. Similar
to the $\ket{W}_N$ state, the Dicke state has the same feature, which is
defined as
\begin{eqnarray}\label{dicke}
\ket{S(N;n)}=\sqrt{\tfrac{n!(N-n)!}{N!}}\sum_{\text{
permutation}}\ket{\underbrace{0\cdots0}_{n}\underbrace{1\cdots1}_{N-n}}.
\end{eqnarray}

\begin{figure}[t]
\center
\includegraphics[bbllx=42, bblly=18, bburx=586, bbury=581,width=5cm]{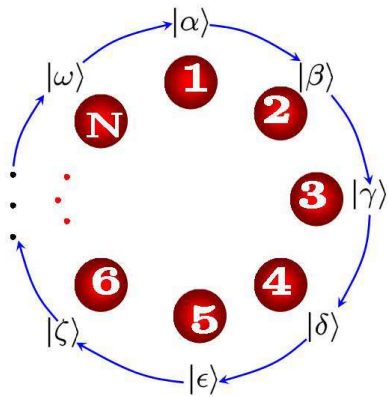}
\caption{\label{1} A schema for the TI of multipartite state, in
which the numbers refer to different parties and the greek letters
denote the single-party states. The arrows represent the cyclic
translation of all single-party states. }
\end{figure}

\emph{Translational invariance} (TI) denotes the situation where the
multipartite state is unchanged under the cyclic translation of the
single-party states. It should pointed out that this definition is
different from the TI defined in physical space, e.g. in lattice
systems, for which TI depends heavily on the geometry of the
physical space. Actually because we are only interested in \emph{the
connectedness of states} belonging to distinguishable parties, it is
unimportant in this case as to how one may realize this multipartite
state in realistic physical systems. The crucial point for the TI of
the multipartite state is that any party could be distinguished from the
others by a definite label, and the single-party state could belong
to any party by a cyclic translation, as show in Fig. \ref{1}.
Admittedly, it is, in general, related to the geometry of the physical
space as to how one should label all parties. However, a sequence of all parties
can always be constructed only if the labels are definite. Our
definition of TI of multipartite states is simply based on this
sequence, and thus is indirect to the geometry of systems. In fact,
the multipartite states studied in this article can be considered
with the geometry of the circle, as shown in Fig. \ref{1}.

An example is the state
\begin{equation}
\ket{\text{GHZ}'}_N=\tfrac{1}{\sqrt{2}}(\ket{1010\cdots10}+\ket{0101\cdots01}),
\end{equation}
which can be obtained by imposing the local unitary operation
$\sigma^x_2\otimes\sigma^x_4\otimes\cdots\otimes\sigma^x_{2n}\otimes\cdots$
on $\ket{\text{GHZ}}_N$. The key feature of $\ket{\text{GHZ}'}_N$ is
the invariance by translating all single-party states cyclicly.
Interestingly, there is a cyclic structure for $\ket{\text{GHZ}'}_N$
where "1" and "0" appear periodically at the next-nearest-neighbor
site \cite{footnote0}. Furthermore this structure also determines
the times of cyclic translation in order to span all terms in the
multipartite state. For
$\ket{\psi^2}_4=\tfrac{1}{2}(\ket{1100}+\ket{0110}+\ket{0011}+\ket{1001})$
as a general example, there is no periodic structure similar to
$\ket{\text{GHZ}'}_N$, and it has to include all possibilities after
cyclic translation for an  arbitrary term in $\ket{\psi^2}_4$ in order
to maintain TI. There is also a hierarchy for the multipartite
states with TI because of the different cyclic structures. Further
discussions will be presented in Secs. \ref{sec: overlapI} and
\ref{sec: overlapII}. A trivial observation is that PI also means
TI, and so the discussion below will not distinguish between them if it is not
necessary.

An important question is what should be the form of the nearest fully
separable state  for an entangled pure state, which decides
the procedure adopted to find $\Lambda_{\max}$. The general method
is to suppose that the nearest fully separable state is still pure
\cite{wg03}, however that has never been proved exactly to the best of our
knowledge. We present a proof for this point here.

\emph{Proof:} Consider the fully separable state defined in
Eq.\eqref{msep} and an entangled pure state $\ket{\psi}$. Then the
overlap is written as
\begin{eqnarray}
\text{Tr}[\rho^{sep}_f\ket{\psi}\bra{\psi}]&=&\sum_i
p_i\text{Tr}[\rho^{(i)}_1\otimes\rho^{(i)}_2\otimes\cdots\rho^{(i)}_N\ket{\psi}\bra{\psi}]\nonumber\\
&=&\sum_i p_i \Delta_i.
\end{eqnarray}
Let $\Delta_1\geq\Delta_2\geq\cdots$, and one has inequality
$\text{Tr}[\rho^{sep}_f\ket{\psi}\bra{\psi}]\leq\Delta_1$. Thus
\begin{equation}
\max_{\{\rho^{sep}_f\}}\{\text{Tr}[\rho^{sep}_f\ket{\psi}\bra{\psi}]\}\Leftrightarrow
\max_{\{\rho^s\}}\text{Tr}[\rho_1\otimes\rho_2\otimes\cdots\otimes\rho_N\ket{\psi}\bra{\psi}],
\end{equation}
where $\rho^s=\rho_1\otimes\rho_2\otimes\cdots\otimes\rho_N$, and
$\Leftrightarrow$ means "equivalent".

The crucial observation is that $\rho^s$ can be rewritten under the
product state basis as
\begin{equation}
\rho^s=\sum_i\varrho_i\ket{\phi_i}\bra{\phi_i},
\end{equation}
where the set of fully separable pure states
$\ket{\phi_i}=\otimes_{n=1}^{N}\ket{\phi_n^{(i)}}$ constitutes a
product state basis. Then
\begin{eqnarray}
\text{Tr}[\rho^s\ket{\psi}\bra{\psi}]=\sum_i\varrho_i\left|\inp{\psi}{\phi_i}\right|^2=\sum_i\varrho_i\delta_i.
\end{eqnarray}
Let $\delta_1\geq\delta_2\geq\cdots$, and then the overlap satisfies
the inequality
\begin{equation}
\text{Tr}[\rho^s\ket{\psi}\bra{\psi}]\leq\delta_1.
\end{equation}
This important result means
\begin{equation}
\max_{\{\rho^s\}}\text{Tr}[\rho^s\ket{\psi}\bra{\psi}]\Leftrightarrow
\max_{\{\ket{\phi}\}}\left|\inp{\psi}{\phi}\right|^2.
\end{equation}
One reaches the final conclusion that the determination of the
maximal overlap with the nearest fully separable state for an entangled
pure state is equivalent to that with a fully separable pure state,
i.e.,
\begin{equation}\label{pm}
\max_{\{\rho^{sep}_f\}}\{\text{Tr}[\rho^{sep}_f\ket{\psi}\bra{\psi}]\}\Leftrightarrow
\max_{\{\ket{\phi}\}}\left|\inp{\psi}{\phi}\right|^2.
\end{equation}

This relation shows that the nearest fully separable state for an
entangled pure state is only \emph{necessarily} pure because some
inequalities appear in this proof. Respecting that $\rho^{sep}_f$
contains only the classical correlation \cite{lm}, GE actually
measures the minimal distance to the classical state in Hilbert
space by an optimal choice of $\rho^{sep}_f$, and thus quantifies
the nonclassical correlation in the entangled state $\ket{\psi}$.

An interesting case is that the maximal $\delta_i$ is not unique,
and then the nearest fully separable state may be mixed. A typical
example is the determination of the nearest fully separable state
for $\ket{\text{GHZ}'}_N$, which is known as the pure state
$\ket{1010\cdots10}$ or $\ket{0101\cdots01}$(see Sec. \ref{sec:
overlapIN}). However, one can check easily that $\Lambda_{\max}$ has
the same value for
$\tfrac{1}{2}(\ket{1010\cdots10}\bra{1010\cdots10}+
\ket{0101\cdots01}\bra{0101\cdots01})$. Thus we claim that it is
enough for the determination of the maximal overlap $\Lambda_{\max}$
for an entangled pure state to focus only on the fully separable
pure state. Thus, in the following sections, we adopt the notation
\begin{eqnarray}\label{fsp}
&&\ket{\phi(a_1, a_2, \cdots, a_N; \theta_1, \theta_2, \cdots,
\theta_N)}\nonumber\\&=&\otimes_{i=1}^N(\sqrt{a_i}\ket{1}_i+\text{e}^{i\theta_i}\sqrt{1-a_i}\ket{0}_i)
\end{eqnarray}
where $a_i\in[0, 1]$ and $\theta_i\in[0, 2\pi)$ denote the
$N$-qubit fully separable pure state (FSPs).

\section{\label{sec: overlapI}finding the Maximal overlap I: role of translational invariance}
With these preparations, we are ready to evaluate $\Lambda_{\max}$
for translationally invariant entangled states. As shown by the
exact evaluations below, there always exists the nearest fully
separable state with the same TI to entangled state, which can be
constructed by the equiprobably incoherent superposition of the FSPs
with the same $\Lambda_{\max}$. This result means that one can reduce
the optimal of $\Lambda_{\max}$ by utilizing the TI of the entangled
state. For this purpose, the discussion in this section is
implemented mainly for the so-called \emph{basic} TI entangled
states with the geometry of  circle, where the meaning of
\emph{basic} is that $\ket{\psi}$ is composed of only one type of
cyclic structure (the hybrid case will be explored in Sec. \ref{sec:
overlapII}). Although  there is no exact proof, some exact or
numerical examples are presented instead in order to demonstrate the
validity of our points.

\subsection{\label{sec: overlapI3}3-qubit case}
It is known that there are two inequivalent multipartite entangled
states in this case: GHZ and $W$ states. Although their GEs have been
studied extensively, the calculation of their $\Lambda_{\max}$ here is to show the general methods for the evaluation of
$\Lambda_{\max}$, adopted in this section.
\newline\\
1. $\ket{\text{GHZ}}=\frac{1}{\sqrt{2}}(\ket{111}+\ket{000})$
\newline\\
The overlap with $\ket{\phi}$ is written as
\begin{eqnarray}
&&\left|\inp{\text{GHZ}}{\phi}\right|^2=\nonumber\\
&&\frac{1}{2}\left|\sqrt{a_1a_2a_3}+\text{e}^{i(\theta_1+\theta_2+\theta_3)}
\sqrt{(1-a_1)(1-a_2)(1-a_3)}\right|^2
\end{eqnarray}
With the inequality
\begin{eqnarray}\label{m1}
x_1^n+x_2^n+\cdots+x_N^n\geq N(x_1x_2\cdots x_N)^{\frac{n}{N}}
\end{eqnarray}
where the equality happens if and only if (iff)
$x_1=x_2=\cdots=x_N$, we have
\begin{eqnarray}
\left|\inp{\text{GHZ}}{\phi}\right|^2&\leq&\frac{1}{2}\left|a^{\frac{3}{2}}+
\text{e}^{i(\theta_1+\theta_2+\theta_3)}(1-a)^{\frac{3}{2}}\right|^2\nonumber\\
&\leq&\frac{1}{2}\left[a^{\frac{3}{2}}+ (1-a)^{\frac{3}{2}}\right]^2
\end{eqnarray}
where the first equality happens iff $a_1=a_2=a_3=a$, and the second
only if $\theta_1+\theta_2+\theta_3=2m\pi (m\text{ is an integer})$.
Consequently the optimal for six independent variables becomes for a
single one, $a$, i.e.,
\begin{equation}
\max_{\ket{\phi}}\left|\inp{\text{GHZ}}{\phi}\right|^2\Leftrightarrow\frac{1}{2}\max_a\left[a^{\frac{3}{2}}+
(1-a)^{\frac{3}{2}}\right]^2,
\end{equation}
which can be shown easily that its maximal values is $\frac{1}{2}$
when $a=1$ or $a=0$.

Then the nearest fully separable state can be chosen as  $\ket{111}$
or $\ket{000}$, or  their incoherent superposition with equal
amplitude $\tfrac{1}{2}(\ket{111}\bra{111}+\ket{000}\bra{000})$.
This observation shows that there always exists the fully separable
state with PI by choosing $\theta_1=\theta_2=\theta_3$, whether it is pure
or mixed. With respect to the limit
$\theta_1+\theta_2+\theta_3=2m\pi$, this result is only necessary.
\newline\\
2. $\ket{W}=\frac{1}{\sqrt{3}}(\ket{100}+\ket{010}+\ket{001})$
\newline\\
By the method discussed in Appendix \ref{aa}, one can easily obtain
$\Lambda_{\max}=4/9$ when $a_1=a_2=a_3=1/3$ and
$\theta_1=\theta_2=\theta_3$. However, in this place we resolve this
question from the point of view of the TI.

It should be emphasized that \emph{TI for a
multipartite state actually defines a relative connection between
the states that belong to different parties.} For example, the GHZ state
implies that all parties would have the same state simultaneously.
This feature leads to the supposition that the nearest fully separable state would be
also permutationally invariant. In contrast, the $W$ state implies that
the nearest neighbor parties always have the same state
\cite{footnote1}, or equivalently that there is always a single party
showing a state that is different from the other two parties. We should
point out that this feature of the $W$ state is more fundamental than PI
(see Sec. \ref{sec: overlapII}), and it is a unique characteristic as compared to
other TI multipartite states. It is thus a natural speculation that
the nearest fully separable state should display the same TI. Thus
we attain the important supposition for the nearest fully separable
state for $\ket{W}$:
\begin{widetext}
\begin{eqnarray}\label{ti}
\rho_f^{W}=&\tfrac{1}{3}&\left(\ket{\phi(a_1, a_2=a_3, \theta_1,
\theta_2=\theta_3)}\bra{\phi(a_1, a_2=a_3, \theta_1,
\theta_2=\theta_3)}\right.\nonumber\\&+&\ket{\phi(a_2, a_1=a_3,
\theta_2, \theta_1=\theta_3)}\bra{\phi(a_2, a_1=a_3, \theta_2,
\theta_1=\theta_3)}\nonumber\\&+&\left.\ket{\phi(a_3, a_1=a_2,
\theta_3, \theta_1=\theta_2)}\bra{\phi(a_3, a_1=a_2, \theta_3,
\theta_1=\theta_2)}\right) ,
\end{eqnarray}
\end{widetext}
where $\ket{\phi}$ and the independent variables $a_n$, $\theta_n$
($n=1,2,3$) are defined in Eq. \eqref{fsp}, and three possible
situations are set to be incoherent so that $\rho_f^{W}$ is
separable. The crucial feature is in $\rho_f^{W}$: that nearest-neighbor parties sharing the same state exist, which is same
as that of $\ket{W}$. Furthermore, because all terms after cyclic
translation are included, $\rho_f^{W}$ is obviously TI.  Next we will
display the correctness of this supposition by obtaining the
same $\Lambda_{\max}$ by calculating the overlap of the $W$ state
and $\rho_f^{W}$.

Similar to the determination of Eq. \eqref{pm}, the key point is to
find the maximal overlaps of the $W$ state and the terms in
$\rho_f^W$. Fortunately they are the same because of the same TI of the $W$
state and $\rho_f^W$, and then it is enough to evaluate the overlap
with one arbitrary term in $\rho_f^{W}$, e.g.,
\begin{eqnarray}
&&\max\text{Tr}\left[\ket{W}\bra{W}\rho_f^W\right]\Leftrightarrow\nonumber\\
&&\frac{1}{3}\max\left|\inp{W}{\phi(a_1, a_2=a_3,
\theta_1, \theta_2=\theta_3)}\right|^2\nonumber\\
&\Rightarrow&\frac{1}{3}\max_{a_1\neq a_2, \theta_1\neq
\theta_2}\left|\text{e}^{2i\theta_2}\sqrt{a_1(1-a_2)^2}+\right.\nonumber\\
&&\left.2\text{e}^{i(\theta_1+\theta_2)}\sqrt{(1-a_1)a_2(1-a_2)}\right|^2\nonumber\\
&\leq&\tfrac{1}{3}\max_{a_1\neq
a_2}\left[\sqrt{a_1(1-a_2)^2}+2\sqrt{(1-a_1)a_2(1-a_2)}\right]^2
\end{eqnarray}
where the equality occurs when $\theta_1=\theta_2$.

The next step is to decide the maximal value of
$\sqrt{a_1(1-a_2)^2}+2\sqrt{(1-a_1)a_2(1-a_2)}$, which can be
obtained easily by calculating its first and second derivation with
independent variables $a_1, a_2$. Thus one can check that the
maximal extremal point occurs at $a_1=a_2=1/3$. Because $\rho_3^{W}$
is TI, the calculation for the other two components in Eq.
\eqref{ti} has the same result. Then the nearest FSPs are
$\rho_3^{W}=\ket{\phi(a_1=a_2=a_3=1/3,
\theta_1=\theta_2=\theta_3)}\bra{\phi(a_1=a_2=a_3=1/3,
\theta_1=\theta_2=\theta_3)}$, which demonstrates the validity of the
supposition $\rho_f^{W}$.

Some comments are in order. The calculations of $\Lambda_{\max}$ for
GHZ and $W$ states have illustrated the general procedure for
determining $\Lambda_{\max}$ for TI entangled states. A crucial
supposition is Eq.\eqref{ti}, which comes directly
from our understanding of the TI in multipartite entangled states,
and can be generalized easily to other multi-qubit states.
Mathematically, in order to find the nearest fully separable state
for the entangled state with a certain symmetry, it is enough to
search into the state subspace with the same symmetry, because the
overlap for the two states in the same subspace are believed to be
no smaller than that in distinct spaces. Particularly, this
supposition is not only valid for a pure state but is valid also for a mixed state,
and a formal statement from the Schwartz-Cauchy inequality can be found
in our recent work \cite{cui10}. In the following discussion, we
should demonstrate the popularity and validity of the supposition
[Eq. \eqref{ti}] through several exact examples.

\subsection{\label{sec: overlapI4}4-qubit case}
There are four basic multipartite entangled states with TI:
$\ket{\text{GHZ}}_4$, $\ket{W}_4$, $\ket{\text{GHZ}'}_4$ and
$\ket{\psi}_4=\tfrac{1}{2}(\ket{1100}+\ket{0110}+\ket{0011}+\ket{1001})$
 \cite{footnote2}. Because a general discussion on
$\ket{\text{GHZ}}_N$, $\ket{W}_N$ and $\ket{\text{GHZ}'}_N$ will be
presented at the end of this section, only $\ket{\psi}_4$ is studied
here. The overlap is determined exactly by the relation
\begin{eqnarray}
&&\left|{_4\inp{\psi}{\phi}}\right|^2\nonumber\\&&=\frac{1}{4}
\left|\text{e}^{i(\theta_3+\theta_4)}\sqrt{a_1a_2(1-a_3)(1-a_4)}+
\text{e}^{i(\theta_1+\theta_4)}\times\right.\nonumber\\
&&\hspace{1em}\sqrt{(1-a_1)a_2a_3(1-a_4)}+\text{e}^{i(\theta_1+\theta_2)}
\sqrt{(1-a_1)}\times\nonumber\\
&&\hspace{1em}\left. \sqrt{(1-a_2)a_3a_4}+
\text{e}^{i(\theta_2+\theta_3)}
\sqrt{a_1(1-a_2)(1-a_3)a_4} \right|^2\nonumber\\
&&=\frac{1}{4}
\left|\left(\text{e}^{i\theta_3}\sqrt{a_1(1-a_3)}+\text{e}^{i\theta_1}\sqrt{(1-a_1)a_3}\right)\times\right.\nonumber\\
&&\hspace{2em}\left.\left(\text{e}^{i\theta_4}\sqrt{a_2(1-a_4)}+\text{e}^{i\theta_2}\sqrt{(1-a_2)a_4}\right)\right|^2
\nonumber\\
&&\leq\frac{1}{4}\left(\sqrt{a_1(1-a_3)}+\sqrt{(1-a_1)a_3}\right)^2\times\nonumber\\
&&\hspace{2em}\left(\sqrt{a_2(1-a_4)}+\sqrt{(1-a_2)a_4}\right)^2,
\end{eqnarray}
where the last equality occurs when $\theta_1=\theta_3$ and
$\theta_2=\theta_4$. It is easy to obtain $\Lambda_{\max}=1/4$
when $a_1+a_3=1$ and $a_2+a_4=1$.

The TI of $\ket{\psi}_4$ is that there are two nearest-neighbor
parties sharing one state, and the other two parties share a
different state. With the above results, one could construct a nearest
fully separable state,
\begin{eqnarray}
\rho_{f}^{\ket{\psi}_4}= \frac{1}{4}\left(\ket{1100}\bra{1100}
+\ket{1001}\bra{1001}+\right.\nonumber\\ \left.
\ket{0011}\bra{0011}+ \ket{0110}\bra{0110}\right)
\end{eqnarray}
with arbitrary $\theta_n (n=1,2,3,4)$ because they cancel each other
out in this case. The above state obviously displays the same TI to
$\ket{\psi}_4$. This result means that one can easily determine
$\Lambda_{\max}$ by utilizing TI of $\ket{\psi}_4$.

An interesting case is the existence of the nearest FSPs with PI
when $\theta_1=\theta_2=\theta_3=\theta_4$ and
$a_1=a_2=a_3=a_4=1/2$. In fact $\ket{\psi}_4$ is biseparable because
$\ket{\psi}_4=\tfrac{1}{\sqrt{2}}(\ket{10}+\ket{01})_{13}\otimes\tfrac{1}{\sqrt{2}}(\ket{10}+\ket{01})_{24}$.
This feature implies that pairs 1-3 and 2-4 are uncorrelated
completely, which forces $\ket{\psi}_4$ into a larger subspace than
that of TI and PI. So the simultaneous existence of the fully
separable state with PI and TI is not surprising in this case.

\subsection{\label{sec: overlapI5}5-qubit case}
Besides $\ket{\text{GHZ}}_5$ and $\ket{W}_5$, there are two
different basic TI entangled states
\begin{eqnarray}
&\ket{\psi^{1a}}_5=\tfrac{1}{\sqrt{5}}\left(\ket{11000}+\ket{01100}+\ket{00110}+\ket{00011}+\ket{10001}\right)
\nonumber\\
&\ket{\psi^{1b}}_5=\tfrac{1}{\sqrt{5}}\left(\ket{10100}+\ket{01010}+\ket{00101}+\ket{10010}+\ket{01001}\right)
\end{eqnarray}
Unfortunately there are no exact results for $\Lambda_{\max}$ of
$\ket{\psi^{1a(b)}}_5$, so one has to rely on numerical
evaluations.

The numerical procedure is to sample exhaustively the possibility of
the values of independent variables as much as possible, and to record
the maximal value of the overlap. Then when the numerical result is not
changed for long sampling times, e.g., $10^4$ or so, this number is
considered to be $\Lambda_{\max}$. For example, the overlap of
$\ket{\psi^{1a}}_5$ is
\begin{eqnarray}
&&\left|{_5\inp{\psi^{1a}}{\phi}}\right|^2\nonumber\\
&=&\frac{1}{5}
\left|\text{e}^{i(\theta_3+\theta_4+\theta_5)}\sqrt{a_1a_2(1-a_3)(1-a_4)(1-a_5)}+\right.\nonumber\\
&&\hspace{1em}\text{e}^{i(\theta_1+\theta_4+\theta_5)}\sqrt{(1-a_1)a_2a_3(1-a_4)(1-a_5)}+\nonumber\\
&&\hspace{1em}\text{e}^{i(\theta_1+\theta_2+\theta_5)}\sqrt{(1-a_1)(1-a_2)a_3a_4(1-a_5)}+\nonumber\\
&&\hspace{1em}\text{e}^{i(\theta_1+\theta_2+\theta_3)}\sqrt{(1-a_1)(1-a_2)(1-a_3)a_4a_5}+\nonumber\\
&&\hspace{1em}\left.\text{e}^{i(\theta_2+\theta_3+\theta_4)}\sqrt{a_1(1-a_2)(1-a_3)(1-a_4)a_5}
\right|^2\nonumber\\
&\leq&\frac{1}{5}
\left[\sqrt{a_1a_2(1-a_3)(1-a_4)(1-a_5)}+\sqrt{(1-a_1)}\times\right.\nonumber\\
&&\hspace{1em}\sqrt{a_2a_3(1-a_4)(1-a_5)}+\sqrt{(1-a_1)(1-a_2)}\times\nonumber\\
&&\hspace{1em}\sqrt{a_3a_4(1-a_5)}+\sqrt{(1-a_1)(1-a_2)(1-a_3)a_4a_5}\nonumber\\
&&\hspace{1em}\left.+\sqrt{a_1(1-a_2)(1-a_3)(1-a_4)a_5} \right]^2
\end{eqnarray}
where the second equality occurs when the values coincide for all
$\theta_{\alpha} (\alpha=1,2,3,4,5)$. Thus the sampling can be
reduced for $a_{\alpha}(\alpha=1,2,3,4,5)$. For reliability, the
sampling times are chosen to be $10^5$ in this section. Finally the value
is steadily closed to $1/5$ and
$\max_{\ket{\phi}}\left|{_5\inp{\psi^{1a}}{\phi}}\right|^2\rightarrow
1/5$.

\begin{figure}[t]
\center
\includegraphics[width=7cm]{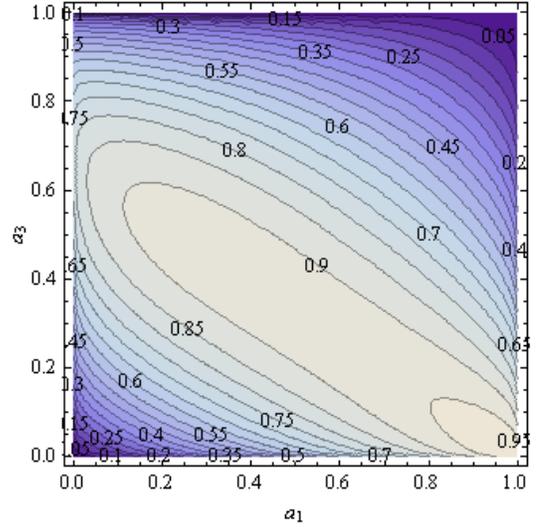}
\caption{\label{p51}(Color online) Contour plot of the function
$f_5^1$ defined in Eq.\eqref{f51}.}
\end{figure}

From the point of view of TI, the main feature of $\ket{\psi^{1a}}_5$ is
that two nearest-neighbor parties have the same state and the other parties
share a different state. Thus an optimal supposition for the nearest
FSPs is, similar to the method discussed in Sec. \ref{sec: overlapI3},
\begin{eqnarray}\label{ti5}
a_1=a_2, a_3=a_4=a_5;\theta_1=\theta_2, \theta_3=\theta_4=\theta_5,
\end{eqnarray}
It should be pointed that this choice is not unique because there are
other different choices, similar to the ones adopted for
Eq.\eqref{ti}. The validity of this supposition can be checked by
asking whether one can obtain the same $\Lambda_{\max}$  by using
the above numerical evaluation. For this purpose, one has
\begin{eqnarray}
&&\left|{_5\inp{\psi^{1a}}{\phi}}\right|^2\nonumber\\
&=&\frac{1}{5}\left|\text{e}^{3i\theta_3}a_1(1-a_3)^{3/2}+2\text{e}^{i(\theta_1+2\theta_3)}
\sqrt{a_1(1-a_1)}\times\right.\nonumber\\
&&\hspace{1em}\left.\sqrt{a_3(1-a_3)^2}+2\text{e}^{i(2\theta_1+\theta_3)}
\sqrt{(1-a_1)^2a_3^2(1-a_3)}\right|^2\nonumber\\
&\leq&\frac{1}{5}\left[a_1(1-a_3)^{3/2}+2
\sqrt{a_1(1-a_1)a_3(1-a_3)^2}\right.\nonumber\\
&&\hspace{1em}\left.+2 \sqrt{(1-a_1)^2a_3^2(1-a_3)}\right]^2
\end{eqnarray}
where the second equality occurs when $\theta_1=\theta_3$. Thus
\begin{eqnarray}\label{f51}
&&\max_{\ket{\phi}}\left|{_5\inp{\psi^{1a}}{\phi}}\right|^2\Rightarrow
\nonumber\\
&&\max_{a_1, a_3}f^1_5=a_1(1-a_3)^{3/2}+2
\sqrt{a_1(1-a_1)a_3(1-a_3)^2}\nonumber\\
&&\hspace{4em}+2 \sqrt{(1-a_1)^2a_3^2(1-a_3)}.
\end{eqnarray}
From Fig. \ref{p51}, $f_5^1$ clearly has the maximal value 1 at
$a_1=1$ and $a_3=0$, and then $\Lambda_{\max}$ is $1/5$ too.

This example displays the validity of the supposition
Eq.\eqref{ti5}. Then the nearest fully separable state with the
same TI to $\ket{\psi^{1a}}_5$ can be constructed readily, as was done
in three- and four-qubit case.  A similar discussion can also be applied
to $\ket{\psi^{1b}}_5$, for which $\Lambda_{\max}$ is also $1/5$.
The difference is that the supposition for nearest FSPs from TI
becomes $a_1=a_3, a_2=a_4=a_5$ and $\theta_1=\theta_3,
\theta_2=\theta_4=\theta_5$ in this case.

\subsection{\label{sec: overlapI6}6-qubit case}
\begin{figure}[t]
\center
\includegraphics[width=7cm]{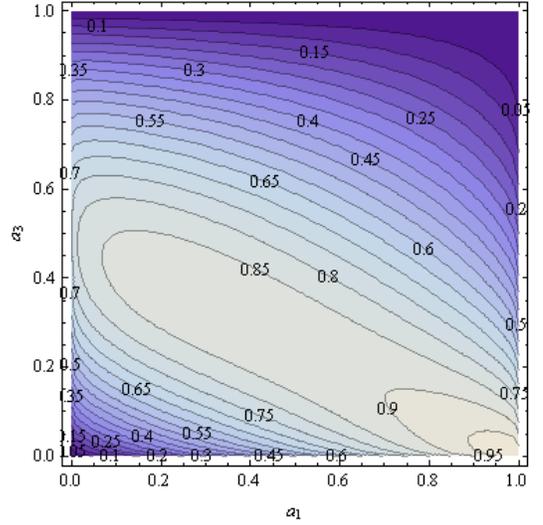}
\caption{\label{p61}(Color online) Contour plot of the function
$f_6^1$ defined in Eq.\eqref{f61}.}
\end{figure}

In order to show further the fundamental role of the TI, the nearest
FSPs and $\Lambda_{\max}$ for six-qubit multipartite entangled states
are studied in this subsection. In addition to $\ket{\text{GHZ}}_6$,
$\ket{W}_6$ and $\ket{\text{GHZ}'}_6$, there are three types of the
basic TI entangled states:
\begin{eqnarray}
\ket{\psi^{1a}}_6&=&\frac{1}{\sqrt{6}}(\ket{110000}+ \text{[the
other cyclic terms]}),\nonumber\\
\ket{\psi^{1b}}_6&=&\frac{1}{\sqrt{6}}(\ket{101000}+ \text{[the
other cyclic terms]}),\nonumber\\
\ket{\psi^{2a}}_6&=&\frac{1}{\sqrt{6}}(\ket{111000}+ \text{[the
other cyclic terms]}),\nonumber\\
\ket{\psi^{2b}}_6&=&\frac{1}{\sqrt{6}}(\ket{101100}+ \text{[the
other cyclic terms]}),\nonumber\\
\ket{\psi^{2c}}_6&=&\frac{1}{\sqrt{6}}(\ket{110100}+ \text{[the
other cyclic terms]}),\nonumber\\
\ket{\psi^3}_6&=&\frac{1}{\sqrt{3}}(\ket{100100}+\ket{010010}+\ket{001001}),
\end{eqnarray}
which will be discussed respectively below.
\newline\\
1. $\ket{\psi^{1a}}_6=\frac{1}{\sqrt{6}}(\ket{110000}+ \text{[the
other cyclic terms]})$
\newline\\
There is no exact result of $\Lambda_{\max}$ for $\ket{\psi^{1a}}$,
and one has to rely on numerical evaluation, as was done in the previous
subsection, which showed $\Lambda_{\max}=1/6$. Now let us repeat the
calculation from the TI point of view. The crucial supposition for the nearest
FSPs is $a_1=a_2, a_3=a_4=a_5=a_6$ and $\theta_1=\theta_2,
\theta_3=\theta_4=\theta_5=\theta_6$, and then
\begin{eqnarray}
&&\left|{_6\inp{\psi^{1a}}{\phi}}\right|^2\nonumber\\
&=&\frac{1}{6}\left|a_1(1-a_3)^2\text{e}^{4i\theta_3}+2\sqrt{a_1(1-a_1)a_3(1-a_3)^3}\times\right.\nonumber\\
&&\hspace{1em}\left.\text{e}^{i(\theta_1+3\theta_3)}+3(1-a_1)a_3(1-a_3)\text{e}^{2i(\theta_1+\theta_3)}\right|^2
\nonumber\\
&\leq&\frac{1}{6}\left[a_1(1-a_3)^2+2\sqrt{a_1(1-a_1)a_3(1-a_3)^3}\right.\nonumber\\
&&\hspace{1em}\left.+3(1-a_1)a_3(1-a_3)\right]^2
\end{eqnarray}
where the second equality occurs when $\theta_1=\theta_3$. Then the
determination of $\Lambda_{\max}$ is reduced to find the maximal
values of
\begin{eqnarray}\label{f61}
f_6^1&=&a_1(1-a_3)^2+2\sqrt{a_1(1-a_1)a_3(1-a_3)^3}\nonumber\\
&&+3(1-a_1)a_3(1-a_3).
\end{eqnarray}
As shown in Fig. \ref{p61}, $f_6^1$ has a maximal value of 1 at $a_1=1,
a_3=0$. Thus $\left|{_6\inp{\psi^{1a}}{\phi}}\right|^2$ has a  maximal
value of $1/6$, which is the same as the numerical result. A similar
study is also applied for $\ket{\psi^{1b}}$ if we suppose $a_1=a_3,
a_2=a_4=a_5=a_6,$ and $\theta_1=\theta_3,
\theta_2=\theta_4=\theta_5=\theta_6$ in this case.

In conclusion, one can still find $\Lambda_{\max}$ efficiently by a TI
of $\ket{\psi^{1a}}_6$
\newline\\
2. $\ket{\psi^{2a}}_6=\frac{1}{\sqrt{6}}(\ket{111000}+ \text{[the
other cyclic terms]})$

After some simplifications, the overlap becomes
\begin{eqnarray}\label{ti62}
&&\left|{_6\inp{\psi^{2a}}{\phi}}\right|^2\leq\nonumber\\
&&\frac{1}{6}\left\{\sqrt{a_3(1-a_6)}\left[\sqrt{a_1(1-a_4)a_2(1-a_5)}+
\sqrt{(1-a_1)}\times\right.\right.\nonumber\\
&&\left.\sqrt{a_4}\left(\sqrt{a_2(1-a_5)}+\sqrt{(1-a_2)a_5}\right)\right]+\sqrt{(1-a_3)a_6}\times\nonumber\\
&&\left[\sqrt{(1-a_1)a_4(1-a_2)a_5}+
\sqrt{a_1(1-a_4)}\left(\sqrt{a_2(1-a_5)}+\right.\right.\nonumber\\
&& \left.\left.\left.\sqrt{(1-a_2)a_5}\right)\right]\right\}^2
\end{eqnarray}
where the equality occurs when the value of
$\theta_{\alpha}(\alpha=1,2,\cdots,6)$ coincides. By applying the
inequalities
\begin{eqnarray}
\sqrt{a_1(1-a_4)}&\leq&\frac{a_1+(1-a_4)}{2},\nonumber\\
\sqrt{a_2(1-a_5)}&\leq&\frac{a_2+(1-a_5)}{2},\nonumber\\
\sqrt{a_3(1-a_6)}&\leq&\frac{a_3+(1-a_6)}{2},
\end{eqnarray}
where the equality occurs iff $a_1+a_4=1, a_2+a_5=1$ and
$a_3+a_6=1$.  Then
\begin{eqnarray}
\left|{_6\inp{\psi^{2a}}{\phi}}\right|^2\Rightarrow\leq\frac{1}{6}\left(1-a_2+a_1a_2+a_2a_3-a_1a_3\right)^2.
\end{eqnarray}
It is easy for the above expression to find the unique extremal
point at $a_1=a_2=a_3=1/2$, and then the extremal value for
$\left|{_6\inp{\psi^{2a}}{\phi}}\right|^2$ is $\tfrac{3}{32}$.
However it is known that the extremal value is not completely equivalent
to the maximal, and so the boundary points when $a_i=1, 0
(i=1,2,\cdots,N)$ have to be checked independently. One can find the
overlap $1/6$ at the boundary point $a_1=a_2=a_3=1, a_4=a_5=a_6=0$,
for example. Because the extremal point is unique, $\Lambda_{\max}$ in
this case is just $1/6$. This exact result is clearly consistent
with the TI structure of $\ket{\psi^{2a}}_6$; there are always  three
nearest-neighbor parties having the same state and the other parties sharing
one different state.
\newline\\
3.
$\ket{\psi^3}_6=\frac{1}{\sqrt{3}}(\ket{100100}+\ket{010010}+\ket{001001})$

The overlap is
\begin{eqnarray}
&&\left|{_6\inp{\psi^{3}}{\phi}}\right|^2\nonumber\\
&=&\frac{1}{3}\left|\text{e}^{i(\theta_2+\theta_3+\theta_5+\theta_6)}\sqrt{a_1(1-a_2)(1-a_3)a_4(1-a_5)}\times\right.\nonumber\\
&&\sqrt{1-a_6}+\text{e}^{i(\theta_1+\theta_3+\theta_4+\theta_6)}\sqrt{(1-a_1)a_2(1-a_3)}\times\nonumber\\
&&\sqrt{(1-a_4)a_5(1-a_6)}+\text{e}^{i(\theta_1+\theta_2+\theta_4+\theta_5)}\sqrt{(1-a_1)}\times\nonumber\\
&&\left.\sqrt{(1-a_2)a_3(1-a_4)(1-a_5)a_6}\right|^2\nonumber\\
&\leq&\frac{1}{3}\left[\sqrt{a_1(1-a_2)(1-a_3)a_4(1-a_5)(1-a_6)}+\right.\nonumber\\
&&\hspace{1em}\sqrt{(1-a_1)a_2(1-a_3)(1-a_4)a_5(1-a_6)}+\nonumber\\
&&\hspace{1em}\left.\sqrt{(1-a_1)(1-a_2)a_3(1-a_4)(1-a_5)a_6}\right]^2\nonumber\\
&=&\frac{1}{3}\left\{\sqrt{(1-a_3)(1-a_6)}\left[\sqrt{a_1a_4(1-a_2)(1-a_5)}\right.\right.\nonumber\\
&&\hspace{1em}\left.+\sqrt{(1-a_1)(1-a_4)a_2a_5}\right]+\sqrt{a_3a_6}\sqrt{(1-a_1)}\times\nonumber\\
&&\hspace{1em}\left.\sqrt{(1-a_4)(1-a_2)(1-a_5)}\right\}^2,
\end{eqnarray}
where the second equality occurs when
$\theta_1+\theta_4=\theta_2+\theta_5=\theta_3+\theta_6$. From the
relations $ab\leq(a^2+b^2)/2$, the above equation becomes
\begin{eqnarray}
\Rightarrow&\leq&\left\{\left[a_1(1-a_2)+(1-a_1)a_2\right](1-a_3)+\right.\nonumber\\
&&\left.(1-a_1)(1-a_2)a_3\right\}^2,
\end{eqnarray}
of which the extremal point appears at $a_1=a_2=a_3=a_0$. In this
case the unique maximal extremal value of
$\left|{_6\inp{\psi^{3}}{\phi}}\right|^2$ is $16/3^5$ for $a_0=1/3$,
while at the boundary  $a_1=1, a_2=a_3=0$ the overlap is
$1/3>16/3^5$. Finally one obtains $\Lambda_{\max}=1/3$ when
$a_1=a_4=1, a_2=a_3=a_5=a_6=0$, for example. With respect to the
flexible limit on $\theta_{\alpha}$, the FPSs can be constructed
readily with the same TI to $\ket{\psi^3}_6$.

\subsection{\label{sec: overlapI8}8-qubit case}
As for seven-qubit case, the discussion is similar to that of the five-qubit
cases. Because there are no exact results for $\lambda_{\max}$ of
seven-qubit entangled states, except for $\ket{\text{GHZ}}_7$ and
$\ket{W}_7$, this situation is ignored in this section and instead
we focus on the eight-qubit case.

The situation becomes complex for eight-qubit multipartite entangled
states because there are several special cyclic structures in this
case. We will focus on these interesting situations because e the exact
results can be obtained.
\newline\\
1. $\ket{\psi^1}_8=\frac{1}{2}\left(\ket{1000\ 1000}+\ket{0100\
0100}+\ket{0010\ 0010}\right.\\ \left.\hspace{5.5em} + \ket{0001\
0001}\right)$

The overlap is
\begin{eqnarray}
&&\left|{_8\inp{\psi^1}{\psi}}\right|^2\nonumber\\
&=&\frac{1}{4}\left|\text{e}^{i\sum_{\alpha\neq
1,5}^8\theta_{\alpha}}\sqrt{a_1a_5}\prod_{\alpha\neq
1,5}^8\sqrt{1-a_{\alpha}}+\text{e}^{i\sum_{\alpha\neq 2,6}^8\theta_{\alpha}}\times\right.\nonumber\\
&& \sqrt{a_2a_6}\prod_{\alpha\neq
2,6}^8\sqrt{1-a_{\alpha}}+\text{e}^{i\sum_{\alpha\neq
3,7}^8\theta_{\alpha}}
\sqrt{a_3a_7}\times\nonumber\\
&&\left.\prod_{\alpha\neq
3,7}^8\sqrt{1-a_{\alpha}}+\text{e}^{i\sum_{\alpha\neq
4,8}^8\theta_{\alpha}} \sqrt{a_4a_8}\prod_{\alpha\neq
4,8}^8\sqrt{1-a_{\alpha}}\right|^2\nonumber\\
&\leq&\frac{1}{4}\left[\sqrt{a_1a_5}\prod_{\alpha\neq
1,5}^8\sqrt{1-a_{\alpha}}+\sqrt{a_2a_6}\prod_{\alpha\neq 2,6}^8\sqrt{1-a_{\alpha}}\right.\nonumber\\
&&\left.+ \sqrt{a_3a_7}\prod_{\alpha\neq
3,7}^8\sqrt{1-a_{\alpha}}+\sqrt{a_4a_8}\prod_{\alpha\neq
4,8}^8\sqrt{1-a_{\alpha}}\right]^2,
\end{eqnarray}
where the second equality occurs when
$\theta_1+\theta_5=\theta_2+\theta_6=\theta_3+\theta_7=\theta_4+\theta_8$.
The above equation can be rewritten as
\begin{eqnarray}
\Rightarrow&\frac{1}{4}&\left\{\left[\sqrt{a_1a_5(1-a_2)(1-a_6)}
+\sqrt{(1-a_1)(1-a_5)}\times\right.\right.\nonumber\\
&&\left.\sqrt{a_2a_6} \right]\sqrt{(1-a_3)(1-a_4)(1-a_7)(1-a_8)}+\nonumber\\
&&\left[\sqrt{a_3a_7(1-a_4)(1-a_8)}
+\sqrt{(1-a_3)(1-a_7)}\times\right.\nonumber\\
&&\left.\left.\sqrt{a_4a_8}
\right]\sqrt{(1-a_1)(1-a_5)(1-a_2)(1-a_6)}\right\}^2\nonumber\\
\leq&\frac{1}{4}&\left\{\left[a_1(1-a_2)+(1-a_1)a_2\right](1-a_3)(1-a_4)+\right.
\nonumber\\&&\left.(1-a_1)(1-a_2)\left[a_3(1-a_4)+(1-a_3)a_4\right]\right\}^2
\end{eqnarray}
where the second equality occurs when $a_1=a_5, a_2=a_6, a_3=a_7$
and $a_4=a_8$ by the inequality $a^2+b^2\geq2|ab|$.

One can check easily that the extremal point of the above equation
appears when $a_1=a_2=a_3=a_4=a_0$, and then the maximal extremal
value is $3^6/4^7$ when $a_0=1/4$. As for the boundary, one can
directly find the overlap $1/4>3^6/4^7$ only if $a_{\alpha}=1$ for
any one of $\alpha=1,2,3,4$, and the other is zero. Then
$\Lambda_{\max}=1/4$. In addition this situation is consistent with
the TI of $\ket{\psi^1}_8$; two single-party states,
four-party separated from each other, always have the same state and
the others share another different state.
\newline\\
2. $\ket{\psi^2}_8=\frac{1}{2}\left(\ket{1100\ 1100}+\ket{0110\
0110}+\ket{0011\ 0011}\right.\\ \left.\hspace{5.5em} + \ket{1001\
1001}\right)$

The overlap is
\begin{eqnarray}
&&\left|{_8\inp{\psi^2}{\psi}}\right|^2\nonumber\\
&=&\frac{1}{4}\left|\text{e}^{i(\theta_3+\theta_4+\theta_7+\theta_8)}\sqrt{a_1a_2
(1-a_3)(1-a_4)a_5a_6}\times\right.\nonumber\\
&&\sqrt{(1-a_7)(1-a_8)}+\text{e}^{i(\theta_1+\theta_4+\theta_5+\theta_8)}\sqrt{(1-a_1)}\times\nonumber\\
&&\sqrt{a_2
a_3(1-a_4)(1-a_5)a_6a_7(1-a_8)}+\nonumber\\
&&\text{e}^{i(\theta_1+\theta_2+\theta_5+\theta_6)}\sqrt{(1-a_1)(1-a_2)
a_3a_4(1-a_5)}\times\nonumber\\
&&\sqrt{(1-a_6)a_7a_8}+\text{e}^{i(\theta_2+\theta_3+\theta_6+\theta_7)}
\sqrt{a_1(1-a_2)}\times\nonumber\\
&&\left.\sqrt{(1-a_3)a_4a_5(1-a_6)(1-a_7)a_8}\right|^2\nonumber\\
&\leq&\frac{1}{4}\left[\sqrt{a_1a_2(1-a_3)(1-a_4)a_5a_6(1-a_7)(1-a_8)}+\right.\nonumber\\
&&\hspace{1em}\sqrt{(1-a_1)a_2a_3(1-a_4)(1-a_5)a_6a_7(1-a_8)}+\nonumber\\
&&\hspace{1em}\sqrt{(1-a_1)(1-a_2)a_3a_4(1-a_5)(1-a_6)a_7a_8}+\nonumber\\
&&\hspace{1em}\left.\sqrt{a_1(1-a_2)(1-a_3)a_4a_5(1-a_6)(1-a_7)a_8}\right]^2
\end{eqnarray}
where the second equality occurs when
$\theta_1+\theta_5=\theta_2+\theta_6=\theta_3+\theta_7=\theta_4+\theta_8$.

The above equation can be rewritten as
\begin{eqnarray}
&\Rightarrow&\tfrac{1}{4}\left[\sqrt{a_1(1-a_3)a_5(1-a_7)}+\sqrt{(1-a_1)a_3(1-a_5)a_7}\right]^2
\nonumber\\&&\times\left[\sqrt{a_2(1-a_4)a_6(1-a_8)}+\sqrt{(1-a_2)a_4(1-a_6)a_8}\right]^2\nonumber\\
&\leq&\tfrac{1}{4}\left[a_1(1-a_3)+(1-a_1)a_3\right]^2\left[a_2(1-a_4)+(1-a_2)a_4\right]^2,
\end{eqnarray}
where the second equality occurs when $a_1=a_5, a_2=a_6,a_3=a_7$
and $a_4=a_8$. The extremal point of the equation above is uniquely
at $a_1=a_2=a_3=a_4=1/2$, at which the maximal extremal value is
$1/2^6$. However one can easily find that the overlap is $1/4$ when
$a_1=1, a_3=0$ and $a_2=1, a_4=0$, for example, which is obviously
larger than $1/2^6$. Thus $\Lambda_{\max}=1/4$ in this case.

The condition $a_1=a_2=1$ and $a_3=a_4=0$ coincides with the TI
structure of $\ket{\psi^2}_8$; there are two pairs of nearest-
neighbor parties, two-party separated from each other, always having
the same state. This observation again displays the underlying
effect of the TI structure of the entangled state on the determination
of the maximal overlap $\Lambda_{\max}$.

Actually $\ket{\psi^2}_8$ is biseparable because
$\ket{\psi^2}_8=\tfrac{1}{\sqrt{2}}(\ket{1010}+\ket{0101})_{1357}
\otimes\tfrac{1}{\sqrt{2}}(\ket{1010}+\ket{0101})_{2468}$. The
above result is also the manifestation of this biseparable
structure.

\subsection{\label{sec: overlapIN}Arbitrary $N$-qubit case: exact results}
For arbitrary $N$-qubit multipartite states, the maximal overlap can
be determined exactly only for some special cases. However, the TI
structure would still play a fundamental role in the evaluation
of $\Lambda_{\max}$, as shown in the following discussions.
\newline\\
1.
$\ket{\psi_{\text{GHZ}}}_N=\sqrt{c}\ket{11\cdots1}+\text{e}^{i\varphi}\sqrt{1-c}\ket{00\cdots0}$,
\newline\\
where $c\in[0,1]$ and $\varphi\in[0, 2\pi)$. It becomes
$\ket{\text{GHZ}}_N$ when $c=1/2$ and $\varphi=0$. The overlap is
\begin{eqnarray}
&&\left|{_N\inp{\psi_{\text{GHZ}}}{\phi}}\right|^2\nonumber\\
&=&\left|\sqrt{c}\prod_{n=1}^N\sqrt{a_n}+\text{e}^{i\left(-\varphi+\sum_{n=1}^N\theta_n\right)}\sqrt{1-c}\prod_{n=1}^N\sqrt{1-a_n}\right|^2
\nonumber\\&\leq&\left[\sqrt{c}\prod_{n=1}^N\sqrt{a_n}+\sqrt{1-c}\prod_{n=1}^N\sqrt{1-a_n}\right]^2\nonumber\\
&\leq&\left[\sqrt{c}a_0^{N/2}+\sqrt{1-c}(1-a_0)^{N/2}\right]^2
\end{eqnarray}
where the second equality occurs when
$\sum_{n=1}^N\theta_n=\varphi$, and the third one occurs when
$a_n=a_0$ for arbitrary $n=1,2,\cdots,N$.

The extremal point of the above equation is determined by the relation
\begin{equation}
\left(\frac{1}{a_0}-1\right)^{\frac{N}{2}-1}=\sqrt{\frac{c}{1-c}},
\end{equation}
which however corresponds to the minimal extremal value. Thus the
maximal overlap can appear only at the boundary points $a_0=1$ or
$a_0=0$ and $\Lambda_{\max}=c \text{ or } (1-c)$, depending on
$c>1-c$ or $c<1-c$. Only if one choose $\theta_n=\varphi/N$ for
arbitrary $n=1,2,\cdots,N$, the resultant fully separable state is
also PI.
\newline\\
2.
$\ket{\psi'_{\text{GHZ}}}_N=\sqrt{c}\ket{1010\cdots10}+\text{e}^{i\varphi}\sqrt{1-c}\ket{0101\cdots01}$
(even $N$)

Obviously $\ket{\psi'_{\text{GHZ}}}_N$ and
$\ket{\psi_{\text{GHZ}}}_N$ can be converted into each other by the local
unitary operation
$\sigma^x_2\otimes\sigma^x_4\otimes\cdots\otimes\sigma^x_{2n}\otimes\cdots$,
and the two states have the same measure of entanglement. However,
the former displays a different TI from the latter because the
operation is only imposed on the even sites \cite{footnote2}.
Generally if two states can be related by local unitary
transformation, it implies only that they have the same measure of
entanglement, but not that the nearest fully separable
state can be determined trivially in the same way.

This state becomes $\ket{\text{GHZ}'}_N$ when $c=1/2$ and
$\varphi=0$. Its overlap is
\begin{eqnarray}
&&\left|{_N\inp{\psi'_{\text{GHZ}}}{\phi}}\right|^2\nonumber\\
&=&\left|\text{e}^{i\sum_{n=1}^{N/2}\theta_{2n}}\sqrt{c}
\prod_{n=1}^{N/2}\sqrt{a_{2n-1}}\prod_{n=1}^{N/2}\sqrt{1-a_{2n}}\right.+\nonumber\\
&&\left.\text{e}^{i\left(-\varphi+\sum_{n=1}^{N/2}\theta_{2n-1}\right)}\sqrt{1-c}
\prod_{n=1}^{N/2}\sqrt{1-a_{2n-1}}\prod_{n=1}^{N/2}\sqrt{a_{2n}}\right|^2\nonumber\\
&\leq&\left[\sqrt{c}
\prod_{n=1}^{N/2}\sqrt{a_{2n-1}}\prod_{n=1}^{N/2}\sqrt{1-a_{2n}}+\sqrt{1-c}\times\right.\nonumber\\
&&\left.
\prod_{n=1}^{N/2}\sqrt{1-a_{2n-1}}\prod_{n=1}^{N/2}\sqrt{a_{2n}}\right]^2\nonumber\\
&\leq&\left[\sqrt{c}a_1^{\frac{N}{4}}(1-a_2)^{\frac{N}{4}}+
\sqrt{1-c}(1-a_1)^{\frac{N}{4}}a_2^{\frac{N}{4}}\right]^2
\end{eqnarray}
where the second equality occurs when
$\sum_{n=1}^{N/2}\left(\theta_{2n-1}-\theta_{2n}\right)=\varphi$,
and the third one occurs when $a_{2n-1}=a_1$ and $a_{2n}=a_2$ for
$n=1,2,\cdots,N/2$ by Eq.\eqref{m1}.

The above equation can be simplified further as
\begin{eqnarray}
\Rightarrow\leq\left[\sqrt{c}a_1^{\frac{N}{2}}+
\sqrt{1-c}(1-a_1)^{\frac{N}{2}}\right]^2,
\end{eqnarray}
where the equality occurs when $a_1=1-a_2$. There is a minimal
extremal point for the above equation , decided by
\begin{equation}
\left(\frac{1}{a_1}-1\right)^{\frac{N}{2}-1}=\sqrt{\frac{c}{1-c}}.
\end{equation}
Thus $\Lambda_{\max}=c \text{ or } (1-c)$, depending on $c>1-c$ or
$c<1-c$, appears when $a_1=1, a_2=0$ or $a_1=0, a_2=1$. Finally,
because the choice for $\theta_{2n}$ and $\theta_{2n-1}$ depends
flexibly on the relation
$\sum_{n=1}^{N/2}\left(\theta_{2n-1}-\theta_{2n}\right)=\varphi$,
one could freely choose $\theta_{2n-1}=\theta_1$,
$\theta_{2n}=\theta_2$, and $\theta_1\neq\theta_2$. Thus one has the
nearest FSPs, which show the same TI structure to
$\ket{\psi'_{\text{GHZ}}}_N$.

Some comments should be made about $\ket{\text{GHZ}'}_N$ because there
are some ambiguities and disputs about its nearest FSPs
\cite{hkwgg09}. The crucial difference between $\ket{\text{GHZ}'}_N$
and $\ket{\psi'_{\text{GHZ}}}_N$ is at the TI; $\ket{\text{GHZ}'}_N$
is TI exactly because $\ket{1010\cdots10}$ and $\ket{0101\cdots01}$
occur with the same probability amplitude, while the basic feature
of $\ket{\psi'_{\text{GHZ}}}_N$ is that the states are the same for the
next-nearest-neighbor parties. Because $\ket{\text{GHZ}'}_N$ is a
special case of $\ket{\psi'_{\text{GHZ}}}_N$, both states then share
this basic feature. Following the discussion above, the common nearest FSPs
are $\ket{1010\cdots10}$ or $\ket{0101\cdots01}$. However, we
emphasize that this result proves exactly the correctness of the
supposition for the nearest fully separable state, based on the TI
structure of $\ket{\text{GHZ}'}_N$ as discussed in Sec. \ref{sec:
overlapI3},
\begin{widetext}
\begin{eqnarray}
\rho_{\text{GHZ'}}^{FSPs}=\frac{1}{2}\left( \ket{\phi(a_{2n+1}=a,
a_{2n}=b, \theta_{2n+1}=\alpha,
\theta_{2n}=\beta)}\bra{\phi(a_{2n+1}=a,
a_{2n}=b, \theta_{2n+1}=\alpha, \theta_{2n}=\beta)}\right.\nonumber\\
\left.+\ket{\phi(a_{2n+1}=b, a_{2n}=a, \theta_{2n+1}=\beta,
\theta_{2n}=\alpha)}\bra{\phi(a_{2n+1}=b, a_{2n}=a,
\theta_{2n+1}=\beta, \theta_{2n}=\alpha)}\right)
\end{eqnarray}
\end{widetext}
Then the nearest fully separable state, from the above exact result,
\begin{equation*}
\frac{1}{2}\left(\ket{1010\cdots10}\bra{1010\cdots10}+
\ket{0101\cdots01}\bra{0101\cdots01}\right),
\end{equation*}
displays the same TI to $\ket{\text{GHZ}'}_N$.
\newline\\
3.
$\ket{W}_N=\tfrac{1}{\sqrt{N}}(\ket{10\cdots0}+\ket{010\cdots0}+\cdots+\ket{0\cdots01})$

The overlap is
\begin{eqnarray}
&&\left|{_N\inp{W}{\phi}}\right|^2\nonumber\\
&=&\frac{1}{N}\left|\text{e}^{i\sum_{n=1}^N\theta_n}\left(\prod_{n=1}^N\sqrt{1-a_n}\right)
\sum_{n=1}^N\sqrt{\frac{a_n}{1-a_n}}\text{e}^{-i\theta_n}\right|^2\nonumber\\
&\leq&\frac{1}{N}\left[\left(\prod_{n=1}^N\sqrt{1-a_n}\right)
\sum_{n=1}^N\sqrt{\frac{a_n}{1-a_n}}\right]^2,
\end{eqnarray}
where the second equality occurs when the values of all $\theta_n$
coincide. Then it is enough to find the maximal values of
\begin{eqnarray}
f_W=\left(\prod_{n=1}^N\sqrt{1-a_n}\right)
\sum_{n=1}^N\sqrt{\frac{a_n}{1-a_n}}.
\end{eqnarray}

First let us find the extremal point for $f_W$, determined by
\begin{eqnarray}
\frac{\partial f_W}{\partial
a_k}=0\Rightarrow\sum_{n=1}^N\sqrt{\frac{a_n}{1-a_n}}=\frac{1}{\sqrt{a_k(1-a_k)}.}
\end{eqnarray}
Because the relation above is satisfied for arbitrary $a_n$,  the
extremal point occurs only if $a_n=a_0$ for arbitrary
$n=1,2,\cdots,N$. Instead, to find the second derivative with $a_n$,
it is convenient to evaluate the maximal value of
\begin{eqnarray}
f'_W= N a_0 (1-a_0)^{N-1}
\end{eqnarray}
which is the overlap for $a_n=a_0$ with $n=1,2,\cdots,N$. It is easy
to find the extremal value of $f'_W$ when $a_0=1/N$. From the second
derivation,
\begin{eqnarray}
\frac{\partial^2 f'_W}{\partial a_0^2}=(N-1)(1-a_0)^{N-1}(Na_0-2),
\end{eqnarray}
thus $a_0=1/N$ is a maximal extremal point and
\begin{equation}
\max f'_W=(1-\frac{1}{N})^{N-1}.
\end{equation}
As for the boundary where the overlap is $1/N$, one can check that
$(1-\frac{1}{N})^{N-1}-\frac{1}{N}\geq0$ for $N\geq 2$.

In conclusion, $\Lambda_{\max}=(1-\frac{1}{N})^{N-1}$ when
$a_n=a_0=1/N$ and $\theta_n=\theta_0$ for arbitrary
$n=1,2,\cdots,N$. Thus the nearest fully separable state is also PI.

\section{\label{sec:PI}periodic structure of translationally invariant entangled state}
It is obvious from previous studies
that the optimal determination of $\Lambda_{\max}$ can be reduced
greatly by utilizing the TI of multipartite entangled states. This
result is based on the observation that the TI of a multipartite
state actually defines the connection between the single-party
states in an unique manner, and, moreover, this connection can be
spread over all single-party states because of TI. It should be
pointed out that  TI is a distinct feature from one to the other state;
for example, GHZ and $W$ states show different TIs by which one can
easily distinguish one from the other. Thus it is conjectured that
a nearest fully separable state would also exist that can
manifest a unique TI of the multipartite state. As shown in this
section, this state can be constructed easily by combining all nearest
FSPs incoherently with the same amplitude. Thus we claim that there
exists a nearest fully separable state with the same TI as that of
an entangled state, though it may be mixed generally.

Focusing on the geometry of the circle as shown in Fig. \ref{1}, there are
diverse cyclic features in the basic TI entangled states studied in
the previous section. For example, the distinct character for
$\ket{\psi_{\text{GHZ}}}_N$ is that all parties share the same
state. From this point of view, depicted in Fig. \ref{1}, it is enough to
translate the whole system only once in order that the state reverted to its original form. Thus we say it is one period, whereas for
$\ket{\text{GHZ}'}_N$ the single-party states for next-nearest-neighbored parties are always the same, and $\ket{\text{GHZ}'}_N$ is
two period because the state has to translate twice back to its original
form, which could be envisioned also from Fig. \ref{1}. This is also
the main reason that $\ket{\text{GHZ}'}_N$ is composed of two terms.
Consequently $\ket{W}_N$ is $N$ period because one has to translate
the system $N$ times in order to span all possible forms. In
addition $\ket{\psi^3}_6$ is three period, and both of
$\ket{\psi^{1(2)}}_8$ are four period.

An important finding from the above discussion is that the period
numbers seem to be the common divisors of the party number $N$.
Besides the trivial common divisors $1$ and $N$, for example there
are two- and three-period TI entangled states for the six-qubit system because
$6=2\times3$. For eight-qubit case, there are two- and four-period TI
entangled state because $8=2\times4$. Because the spatial of the state
is a circle, it is not strange that TI is heavily dependent on the
number of party.

In conclusion, \emph{for a ($N=n\times m$)-qubit system, there always
exist $n$- and $m$-period TI entangled states.} Thus by exploring
the common divisors for $N$, one can find all possible cyclic
structures of the $N$-qubit TI entangled state. As for prime $N$, there
are only trivial one- and $N$-period states, as shown for the five-qubit
system. Furthermore, we point out that \emph{one can construct more
complicated TI entangled states principally by combining the basic
TI entangled states with different periodic structures}; for example,
the Dicke state for the four-qubit system $\ket{S(4;2)}$, defined in
Eq. \eqref{dicke}, can be decomposed into
$\ket{S(4;2)}=\tfrac{1}{\sqrt{3}}\ket{\text{GHZ}'}_4+\sqrt{\tfrac{2}{3}}\ket{\psi}_4$.
It is the  reason that we call them basic TI entangled states, which we studied in
previous section.

An entangled state composed of different TI structures is called
a hybrid. Then an important question is how to find the nearest fully
separable state for the hybrid TI entangled states. And it is more
interesting whether $\Lambda_{\max}$ is related to the different TIs
in this state. If the answer is yes, it would simplify greatly the
determination of $\Lambda_{\max}$. In the next section, we would
present an explicit study of this question.

\section{\label{sec: overlapII}Finding the maximal overlap II: hybrid Translational invariance and the hierarchy of the periodic structure}
For hybrid TI entangled states, $\Lambda_{\max}$ is in general
difficult to evaluate analytically. By presenting several numerical
examples in this section, we try to illustrate some general features
for these types of  states, which are helpful in reducing the optimal
determination for $\Lambda_{\max}$ in this case.

Unfortunately, we cannot provide an exact result to solidify the
conclusions summarized at the end of this section, and instead have
to rely on the numerical evaluation. For simplicity, the examples
below will focus on the hybrid TI entangled states (HTIEs), composed
of only two basic entangled states with different TIs. The goal is
to find the effect of TI in HTIEs on the optimal determinations of
$\Lambda_{\max}$ and the nearest FSPs. In addition, a hierarchy for
the basic TI entangled states can be found.

The numerical procedure is stated below. Because the HTIEs in this
section includes only two different TI structures, it is natural to
suppose that the nearest FSPs would be determined by one of the two
periodic structures. This supposition comes directly from the
discussion in Sec.\ref{sec: overlapI}. However the supposed
nearest fully separable states have to include two TIs.  Then two
numerical evaluations, called \emph{case 1} and \emph{case 2},
are implemented respectively for two optimally supposed nearest FSPs
based on the TIs in HTIEs. The other two situations, called
\emph{case 0} and \emph{case 3}, correspond respectively to no
supposition and the PI supposition for the
nearest FSPs in order to check the validity of the two former 
suppositions.

The numerical procedure is the same as discussed in Sec. \ref{sec:
overlapI5}; we sample randomly the permissible values of all variables
as much as possible and find the maximal overlap, which is unchanged
for $10^4$ sampling times or so. The critical problem is to justify
which supposition is optimal. Our method is to find one of the
suppositions, for which the numerical result of the overlap is
maximal after the same sampling times. Then we claim that this case
is just the optimal requirement for the nearest FSPs in order to
determine $\Lambda_{\max}$. The reason is simple; it is known that
the random sampling evaluation could attain the same value
theoretically only if the sampling times are infinite. With finite
sampling times, the optimal one is believed to attain the maximal
value more quickly than the nonoptimal sampling time. The sampling times are
$10^5$ in order to guarantee the reliability of the numerical result
and to control the time of evaluation. We should emphasize that this
numerical evaluation is to show the effect of TI of multipartite
entangled states on the determinations of FSPs and  $\Lambda_{\max}$
rather than finding the exact results, because the sampling times are
far from exhaustive. More details of the calculation are shown with
special examples.

\subsection{3-qubit case}
For the three-qubit system, there are only two basic TI entangled states:
$\ket{\text{GHZ}}$ of one-period and $\ket{W}$ of three-period. It is
known that the nearest fully separable states for the two states are
both permutationally invariant. Thus for their HTIEs, the
nearest FSPs are definitely permutationally invariant also. So this
case is omitted in this section.

\subsection{\label{sec: overlapII4}4-qubit case}
There are one, two and four-period basic TI entangled states in this case,
$\ket{\text{GHZ}}_4$, $\ket{\text{GHZ}'}_4$ and $\ket{W}_4$,
$\ket{\psi}_4$. Then there are five different HTIEs:
\begin{eqnarray}\label{m4q}
\text{A1: }
\ket{\psi}_4^{\text{A1}}&=&\sqrt{c}\ket{\text{GHZ}}_4+\text{e}^{i\varphi}\sqrt{1-c}\ket{\psi}_4,
\nonumber\\
\text{A2: }
\ket{\psi}_4^{\text{A2}}&=&\sqrt{c}\ket{\text{GHZ}}_4+\text{e}^{i\varphi}\sqrt{1-c}\ket{\text{GHZ}'}_4,
\nonumber\\
\text{A3: }
\ket{\psi}_4^{\text{A3}}&=&\sqrt{c}\ket{W}_4+\text{e}^{i\varphi}\sqrt{1-c}\ket{\psi}_4,
\nonumber\\
\text{A4: }
\ket{\psi}_4^{\text{A4}}&=&\sqrt{c}\ket{W}_4+\text{e}^{i\varphi}\sqrt{1-c}\ket{\text{GHZ}'}_4,
\nonumber\\
\text{A5: }
\ket{\psi}_4^{\text{A5}}&=&\sqrt{c}\ket{\text{GHZ}'}_4+\text{e}^{i\varphi}\sqrt{1-c}\ket{\psi}_4,
\end{eqnarray}
where $c\in[0,1]$ and $\varphi\in[0, 2\pi)$.
$\varphi$ is set to be $\pi/3$ for all numerical evaluations below.

\begin{table}
\begin{tabular}{p{1.5cm} p{1.5cm} p{1.5cm} p{1.5cm} r}
\hline\hline\text{State}   &  Case 0  &  Case 1  &  Case 2  &  Case 3 \\
\hline
A1-1& 0.28253& - & 0.29458 & {\bf 0.2953} \\
A1-2& 0.28452& - & 0.29077 & {\bf 0.2916} \\
A1-3& 0.34184& - &0.37407 &{\bf 3/8=0.375} \\ \hline
A2-1&0.3495&-&{\bf3/8=0.375}&0.18\\
A2-2&0.24912&-&{\bf 0.25}&{\bf 0.25}\\
A2-3&0.33599&-&0.37282&{\bf3/8=0.375}\\ \hline
A3-1&0.47811&-&0.50217&{\bf0.50364}\\
A3-2&0.51147&-&0.57522&{\bf0.5782}\\
A3-3&0.52825&-&0.58671&{\bf0.58832}\\ \hline
A4-1&0.44235&-&{\bf0.49037}&0.35\\
A4-2&0.43111&-&{\bf0.48201}&0.444\\
A4-3&0.47361&-&0.49538&{\bf0.49580}\\ \hline
A5-1&0.29404&0.29524&0.29518&{\bf0.29530}\\
A5-2&0.28207&{\bf0.29157}&0.27564&0.27589\\
A5-3&0.36237&{\bf3/8=0.375}&0.23274&0.2328\\ \hline\hline
\end{tabular}
\caption{\label{m4}Numerical results of the maximal overlap for
four-qubit mixed TI entangled states defined in Eqs.\eqref{m4q} with
$\varphi=\tfrac{\pi}{3}$.}
\end{table}

In order to highlight the effect of TIs in HTIEs, $c$ is set to be
three different values in this section, $c=1/4, 1/2, 3/4$, which
are labeled as A$N$-1, A$N$-2 and
A$N$-3, as shown in Table \ref{m4}. $N=1,2,3,4,5$ denote
the different HTIEs in Eq.\eqref{m4q}, and the letter A is used
to distinguish them from the other multi-qubit cases. Similar notations
are also applied for the remainder of this section. The different
choices for the values of $c$ are to obtain a general
conclusion that is independent on the special superposition coefficients.
The notations \emph{case 1} and \emph{case 2} in Table \ref{m4},
respectively denote the two suppositions for the nearest FSPs, based
on the TIs of the first and second term in HTIEs Eq. \eqref{m4q}. For
example, for $\ket{\psi}_4^{\text{A5}}$ \emph{case 1} means the
supposition for the nearest FSPs $a_1=a_3, a_2=a_4$ and
$\theta_1=\theta_3, \theta_2=\theta_4$, based on the two-period TI of
$\ket{\text{GHZ}'}_4$, while \emph{case 2} means the supposition
$a_1=a_2, a_3=a_4$ and $\theta_1=\theta_2, \theta_3=\theta_4$, based
on the four-period TI of $\ket{\psi}_4$.

The numerical results are shown in Table \ref{m4}. Interestingly a
hierarchy for the basic TI entangled states can be found. For the
state $\ket{\psi}_4^{\text{A1(3)}}$, the maximal overlap occurs
with the supposition of permutation invariance for the nearest FSPs.
This feature is not surprising because the nearest FSPs for
$\ket{\psi}_4$ may be permutationally invariant, as shown in
Sec.\ref{sec: overlapI4}. In contrast, the maximal overlap for
$\ket{\psi}_4^{\text{A5}}$ can be attained only for FSPs decided by
$\ket{\text{GHZ}'}_4$, as shown by A5-2 and A5-3 in Table \ref{m4}.
As for A5-1, the difference between case 1 and case 3 is just of
order $10^{-5}$, which can be seen as the numerical error. This
observation implies that the TI structure of $\ket{\text{GHZ}'}_4$
is predominant for the determination of $\Lambda_{\max}$ for
$\ket{\psi}_4^{\text{A5}}$, and thus we say $\ket{\text{GHZ}'}_4$
has a higher order of symmetry than $\ket{\psi}_4$, i.e.
$\mathcal{S}_{\ket{\text{GHZ}'}_4 }> \mathcal{S}_{\ket{\psi}_4}$.
For $\ket{\psi}_4^{\text{A2}}$, the numerical results for A2 in
Table \ref{m4} show that $\Lambda_{\max}$ depends only on the
superposition coefficients, which imply that $\ket{\text{GHZ}}_4$
and $\ket{\text{GHZ}'}_4$ have the same order of symmetry,
$\mathcal{S}_{\ket{\text{GHZ}}_4}\sim\mathcal{S}_{\ket{\text{GHZ}'}_4}$.
As for A4, one can obtain the order
$\mathcal{S}_{\ket{\text{GHZ}'}_4}>\mathcal{S}_{\ket{W}}$.

It is intricate for the relation of $\ket{W}$ and $\ket{\psi}_4$;
only by A3 in Table \ref{m4}, we cannot tell which has a higher order
of symmetry because both of the nearest FSPs for $\ket{W}$ and
$\ket{\psi}_4$ are permutationally invariant. However, we should
point out that $\ket{\psi}_4$ actually is biseparable, which is the main
reason for the existence of the PI nearest FSPs. Thus we set
$\mathcal{S}_{\ket{W}_4}>\mathcal{S}_{\ket{\psi}_4}$.

In conclusion, one has
\begin{equation}
\mathcal{S}_{\ket{\text{GHZ}}_4}\sim\mathcal{S}_{\ket{\text{GHZ}'}_4}>\mathcal{S}_{\ket{W}_4}>\mathcal{S}_{\ket{\psi}_4}.
\end{equation}

\subsection{\label{sec: overlapII5}5-qubit case}
\begin{table}
\begin{tabular}{p{1.5cm} p{1.5cm} p{1.5cm} p{1.5cm} r}
\hline\hline\text{State}   &  Case 0  &  Case 1  &  Case 2  &  Case 3 \\
\hline
B1-1& 0.20702& - & 0.24871 & \emph{\bf 0.25124} \\
B1-2& 0.24403& - & 0.27173 & \emph{\bf 0.27329} \\
B1-3& 0.31523& - &0.37251 &\emph{\bf 3/8=0.375} \\ \hline
B2-1&0.33628&-&0.40932&\emph{\bf0.4102}\\
B2-2&0.44841&-& 0.49339&\emph{\bf 0.49597}\\
B2-3&0.39626&-&0.52329&\emph{\bf0.52734}\\ \hline
B3-1&0.22777&0.24746&0.24749&\emph{\bf0.24763}\\
B3-2&0.22326&0.25915& 0.25899&\emph{\bf 0.25920}\\
B3-3&0.22985&0.24752&0.24707&\emph{\bf0.24762}\\ \hline\hline
\end{tabular}
\caption{\label{m5}Numerical results of the maximal overlap for
five-qubit mixed TI entangled states defined in Eqs.\eqref{m5q} with
$\varphi=\tfrac{\pi}{3}$.}
\end{table}

This situation is simple because $5$ is prime number, and there are only one-
and five-period TI entangled states. Then the HTIEs are
\begin{eqnarray}\label{m5q}
\text{B1:
}\ket{\psi}_5^{\text{B1}}&=&\sqrt{c}\ket{\text{GHZ}}_5+\text{e}^{i\varphi}\sqrt{1-c}\ket{\psi^{1a}}_5,
\nonumber\\
\text{B2: }
\ket{\psi}_5^{\text{B2}}&=&\sqrt{c}\ket{W}_5+\text{e}^{i\varphi}\sqrt{1-c}\ket{\psi^{1a}}_5,\nonumber\\
\text{B3: }
\ket{\psi}_5^{\text{B3}}&=&\sqrt{c}\sqrt{1-c}\ket{\psi^{1a}}_5+\text{e}^{i\varphi}\sqrt{1-c}\ket{\psi^{1b}}_5.
\end{eqnarray}
As shown in Table \ref{m5}, the optimal fully separable states can
be obtained always under the supposition of PI. Thus one has the order of symmetry
$\mathcal{S}_{\ket{\text{GHZ}}_5}>\mathcal{S}_{\ket{\psi^{1a}}_5}$
and $\mathcal{S}_{\ket{W}_5}>\mathcal{S}_{\ket{\psi^{1a}}_5}$. As
for $\ket{W}_5$ and $\ket{\text{GHZ}}_5$, we could temporarily set
$\mathcal{S}_{\ket{\text{GHZ}}_5}>\mathcal{S}_{\ket{W}_5}$ because
$\ket{\text{GHZ}}_5$ is one-period, while $\ket{W}_5$ is five-period. A
strange result of B3 is that the optimal FSPs is permutationally
invariant, adn independent on the TI structures of components in
$\ket{\psi}_5^{\text{B3}}$. Thus $\ket{\psi^{1a}}_5$ and
$\ket{\psi^{1b}}_5$ could be considered same order.

In total one has a hierarchy of the symmetry, in this case
\begin{equation}
\mathcal{S}_{\ket{\text{GHZ}}_5}>\mathcal{S}_{\ket{W}_5}>\mathcal{S}_{\ket{\psi^{1a}}_5}\sim\mathcal{S}_{\ket{\psi^{1b}}_5}.
\end{equation}

\subsection{\label{sec: overlapII6}6-qubit case}
\begin{table}
\begin{tabular}{p{1.5cm} p{1.5cm} p{1.5cm} p{1.5cm} r}
\hline\hline\text{State}   &  Case 0  &  Case 1  &  Case 2  &  Case 3 \\
\hline
C1-1& 0.24839& - & {\bf 3/8=0.375} & 1/8=0.125 \\
C1-2& 0.18059& - & {\bf 0.25} & {\bf 0.25} \\
C1-3& 0.25466& - &0.37159 &{\bf 3/8=0.375} \\ \hline
C2-1&0.29698&-&{\bf 3/8=0.375}&0.159\\
C2-2&0.21546&-&{\bf0.2672}& 0.25793\\
C2-3&0.25309&-&{\bf 0.34657}&{\bf0.34658}\\ \hline
C3-1&0.16963&-&0.1952&{\bf0.20225}\\
C3-2&0.20256&-& 0.24579&{\bf 0.25706}\\
C3-3&0.2557&-&0.35873&{\bf3/8=0.375}\\ \hline
C4-1&0.24022&-&0.3537&{\bf0.35595}\\
C4-2&0.30666&-& 0.44431&{\bf 0.44687}\\
C4-3&0.33181&-&0.48512&{\bf0.48992}\\ \hline
C5-1&0.11394&-&0.12491&{\bf1/8=0.125}\\
C5-2&0.19292&-& 0.24775&{\bf 0.25}\\
C5-3&0.2892&-&0.37285&{\bf3/8=0.375}\\ \hline
C6-1&0.17516&-&0.22896&{\bf0.22998}\\
C6-2&0.20439&-& 0.31553&{\bf 0.31641}\\
C6-3&0.24171&-&0.38406&{\bf0.38689}\\ \hline
C7-1&0.1474&0.15244&0.15161&{\bf0.153}\\
C7-2&0.15756&{\bf 0.25}& 0.1344&0.135\\
C7-3&0.2534&{\bf 3/8=0.375}&0.10211&0.10273\\ \hline
C8-1&0.11912&{\bf 1/8=0.125}&0.12456&0.1015\\
C8-2&0.21251&{\bf 0.25}& 0.08947&0.09\\
C8-3&0.27651&{\bf 3/8=0.375}&0.07025&0.07\\ \hline
C9-1&0.19456&-&{\bf 0.25}&0.14\\
C9-2&0.19164&-& 0.24937&{\bf 0.25}\\
C9-3&0.27267&-&0.37207&{\bf 3/8=0.375}\\ \hline
C10-1&0.25579&-&{\bf 0.33068}&0.2603\\
C10-2&0.26902&-& {\bf 0.36098}&{\bf 0.36095}\\
C10-3&0.26567&-&{\bf 0.42781}&{\bf 0.42782}\\ \hline
C11-1&0.18722&0.12411&{\bf 0.25}&0.09\\
C11-2&0.18246&{\bf 0.25}& 0.16624&0.085\\
C11-3&0.25725&{\bf 3/8=0.375}&0.08286&0.072\\ \hline
C12-1&0.13548&0.15535&{\bf 0.15547}&{\bf 0.15555}\\
C12-2&0.13392&{\bf 1/6$\approx$0.1667}& 0.14518&0.14532\\
C12-3&0.20826&{\bf 0.25}&0.12247&0.12263\\ \hline
C13-1&0.12497&{\bf 0.14731}&{\bf 0.14731}&{\bf 0.14731}\\
C13-2&0.11305&{\bf 1/6$\approx$0.1667}& 0.14542&0.14542\\
C13-3&0.15592&{\bf 0.25}&0.12959&0.12960\\ \hline\hline
\end{tabular}
\caption{\label{m6}Numerical results of the maximal overlap for
six-qubit mixed TI entangled states defined in Eqs.\eqref{m6q} with
$\varphi=\tfrac{\pi}{3}$.}
\end{table}

This case is more complicated because there are four types of basic
TI entangled states. Thus the discussion focuses on the
superposition of basic TI entangled states with different periodic
structures,
\begin{eqnarray}\label{m6q}
\text{C1:
}\ket{\psi}_6^{\text{C1}}&=&\sqrt{c}\ket{\text{GHZ}}_6+\text{e}^{i\varphi}\sqrt{1-c}\ket{\text{GHZ}'}_6,
\nonumber\\
\text{C2: }
\ket{\psi}_6^{\text{C2}}&=&\sqrt{c}\ket{W}_6+\text{e}^{i\varphi}\sqrt{1-c}\ket{\text{GHZ}'}_6,
\nonumber\\
\text{C3: }
\ket{\psi}_6^{\text{C3}}&=&\sqrt{c}\ket{\text{GHZ}}_6+\text{e}^{i\varphi}\sqrt{1-c}\ket{\psi^{1a}}_6,
\nonumber\\
\text{C4: }
\ket{\psi}_6^{\text{C4}}&=&\sqrt{c}\ket{W}_6+\text{e}^{i\varphi}\sqrt{1-c}\ket{\psi^{1a}}_6,
\nonumber\\
\text{C5: }
\ket{\psi}_6^{\text{C5}}&=&\sqrt{c}\ket{\text{GHZ}}_6+\text{e}^{i\varphi}\sqrt{1-c}\ket{\psi^{2a}}_6,
\nonumber\\
\text{C6: }
\ket{\psi}_6^{\text{C6}}&=&\sqrt{c}\ket{W}_6+\text{e}^{i\varphi}\sqrt{1-c}\ket{\psi^{2a}}_6,
\nonumber\\
\text{C7: }
\ket{\psi}_6^{\text{C7}}&=&\sqrt{c}\ket{\text{GHZ}'}_6+\text{e}^{i\varphi}\sqrt{1-c}\ket{\psi^{1a}}_6,
\nonumber\\
\text{C8: }
\ket{\psi}_6^{\text{C8}}&=&\sqrt{c}\ket{\text{GHZ}'}_6+\text{e}^{i\varphi}\sqrt{1-c}\ket{\psi^{2a}}_6,
\nonumber\\
\text{C9: }
\ket{\psi}_6^{\text{C9}}&=&\sqrt{c}\ket{\text{GHZ}}_6+\text{e}^{i\varphi}\sqrt{1-c}\ket{\psi^{3}}_6,
\nonumber\\
\text{C10: }
\ket{\psi}_6^{\text{C10}}&=&\sqrt{c}\ket{W}_6+\text{e}^{i\varphi}\sqrt{1-c}\ket{\psi^{3}}_6,
\nonumber\\
\text{C11: }
\ket{\psi}_6^{\text{C11}}&=&\sqrt{c}\ket{\text{GHZ}'}_6+\text{e}^{i\varphi}\sqrt{1-c}\ket{\psi^{3}}_6,
\nonumber\\
\text{C12: }
\ket{\psi}_6^{\text{C12}}&=&\sqrt{c}\ket{\psi^{3}}_6+\text{e}^{i\varphi}\sqrt{1-c}\ket{\psi^{1a}}_6,
\nonumber\\
\text{C13: }
\ket{\psi}_6^{\text{C13}}&=&\sqrt{c}\ket{\psi^{3}}_6+\text{e}^{i\varphi}\sqrt{1-c}\ket{\psi^{2a}}_6.
\end{eqnarray}
$\ket{\psi^{1a}}_6$ and $\ket{\psi^{1b}}_6$ are considered
equivalent numerically because the only difference is the position of
single-party state $\ket{1}$s, therefore $\ket{\psi^{2a(b,c)}}_6$.

The numerical results are listed in Table \ref{m6}. Some features
similar to those of the four-qubit case can be found. It is obvious from C1 in
Table \ref{m6} that $\Lambda_{\max}$ depends only on superposition
coefficients, and thus
$\mathcal{S}_{\ket{\text{GHZ}}_6}\sim\mathcal{S}_{\ket{\text{GHZ}'}_6}$.
For C2, the supposition for the nearest fully separable state is
optimal from the TI structure of $\ket{\text{GHZ}'}_6$, and thus
$\mathcal{S}_{\ket{\text{GHZ}'}_6}>\mathcal{S}_{\ket{W}_6}$. By C3,
C4, C5, C6, C7, and C8 in Table \ref{m6}, $\ket{\psi^{1a}}_6$ and
$\ket{\psi^{2a}}_6$ both show lower order of the symmetry than
$\ket{\text{GHZ}}_6$, $\ket{\text{GHZ}'}_6$ and $\ket{W}_6$ because
$\Lambda_{\max}$ is decided mainly by the TI of the latter.

A particular state in this case is $\ket{\psi^{3}}_6$, which is
three-period. By C9 and C11 in Table \ref{m6}, $\Lambda_{\max}$ depends
only on the superposition coefficients. However, because
$\ket{\psi^{3}}_6$ is three-period, thus
$\mathcal{S}_{\ket{\psi^{3}}_6}<\mathcal{S}_{\ket{\text{GHZ}}_6}\sim\mathcal{S}_{\ket{\text{GHZ}'}_6}$,
whereas from C10, C12, and C13, $\ket{\psi^{3}}_6$ always has
a noticeable effect on $\Lambda_{\max}$.

In conclusion, one obtains
\begin{eqnarray}
\mathcal{S}_{\ket{\text{GHZ}}_6}\sim\mathcal{S}_{\ket{\text{GHZ}'}_6}>\mathcal{S}_{\ket{\psi^{3}}_6}>\mathcal{S}_{\ket{W}_6}
>\mathcal{S}_{\ket{\psi^{1(2)a}}_6}
\end{eqnarray}
It is exceptional for the relation of $\ket{\psi^{1a}}_6$ and
$\ket{\psi^{2a}}_6$, as will shown at the end of this section.

\subsection{\label{sec: overlapII8}8-qubit case}
\begin{table}
\begin{tabular}{p{1.5cm} p{1.5cm} p{1.5cm} p{1.5cm} r}
\hline\hline\text{State}   &  Case 0  &  Case 1  &  Case 2  &  Case 3 \\
\hline
D1-1& 0.17958& - & \emph{\bf 3/8=0.375} & 1/8=0.125 \\
D1-2& 0.11301& - & \emph{\bf 0.25} & \emph{\bf 0.25} \\
D1-3& 0.16164& - &0.37167 &\emph{\bf 3/8=0.375} \\ \hline
D2-1&0.16865&-&\emph{\bf 3/8=0.375}&0.12\\
D2-2&0.1515&-& \emph{\bf0.25}& 0.20792\\
D2-3&0.14598&-&{\bf 0.30415}&\emph{\bf0.30416}\\ \hline
D3-1&0.09073&-&{\bf 3/16=0.1875}&0.1307\\
D3-2&0.12702&-& 0.24895&\emph{\bf 0.25}\\
D3-3&0.14792&-&0.37068&\emph{\bf3/8=0.375}\\ \hline
D4-1&0.12409&-&{\bf 0.24924}&0.22288\\
D4-2&0.17166&-& {\bf 0.32373}&\emph{\bf 0.32373}\\
D4-3&0.15625&-&{\bf0.39656}&\emph{\bf0.39656}\\ \hline
D5-1&0.08002&-&{\bf 3/16=0.1875}&1/8=0.125\\
D5-2&0.13844&-& 0.24804&\emph{\bf 0.25}\\
D5-3&0.14722&-&0.3733&\emph{\bf3/8=0.375}\\ \hline
D6-1&0.08814&-&{\bf3/16$\approx$0.187}0.185&0.11464\\
D6-2&0.09481&-& {\bf0.21338}&\emph{\bf 0.21337}\\
D6-3&0.15491&-&0.30844&\emph{\bf0.30846}\\ \hline
D7-1&0.11729&0.13375&{\bf3/16=0.1875}&0.04402\\
D7-2&0.11472&{\bf 0.25}&0.12349& 0.03601\\
D7-3&0.2053&{\bf 3/8=0.375}&0.06219&0.02544\\ \hline
D8-1&0.0962&0.12346&{\bf3/16=0.1875}&0.1015\\
D8-2&0.13414&{\bf 0.25}& 0.12402&0.017\\
D8-3&0.17557&{\bf 3/8=0.375}&0.06204&0.01455\\ \hline
D9-1&0.07897&{\bf3/16=0.1875}&0.06424&0.3486\\
D9-2&0.07395&{\bf 1/8}& {\bf 1/8}&0.04389\\
D9-3&0.09303&0.0258&{\bf3/16=0.1875}&0.04943\\ \hline\hline
\end{tabular}
\caption{\label{m8}Numerical results of the maximal overlap for
eight-qubit mixed TI entangled states defined in Eqs.\eqref{m8q}.}
\end{table}

Eight-qubit HTIEs are studied in order to display the generality of the
hierarchy for basic TI entangled states,
\begin{eqnarray}\label{m8q}
\text{D1:
}\ket{\psi}_6^{\text{D1}}&=&\sqrt{c}\ket{\text{GHZ}}_8+\text{e}^{i\varphi}\sqrt{1-c}\ket{\text{GHZ}'}_8,
\nonumber\\
\text{D2: }
\ket{\psi}_6^{\text{D2}}&=&\sqrt{c}\ket{W}_8+\text{e}^{i\varphi}\sqrt{1-c}\ket{\text{GHZ}'}_8,
\nonumber\\
\text{D3: }
\ket{\psi}_6^{\text{D3}}&=&\sqrt{c}\ket{\text{GHZ}}_8+\text{e}^{i\varphi}\sqrt{1-c}\ket{\psi^{1}}_8,
\nonumber\\
\text{D4: }
\ket{\psi}_6^{\text{D4}}&=&\sqrt{c}\ket{W}_8+\text{e}^{i\varphi}\sqrt{1-c}\ket{\psi^{1}}_8,
\nonumber\\
\text{D5: }
\ket{\psi}_6^{\text{D5}}&=&\sqrt{c}\ket{\text{GHZ}}_8+\text{e}^{i\varphi}\sqrt{1-c}\ket{\psi^{2}}_8,
\nonumber\\
\text{D6: }
\ket{\psi}_6^{\text{D6}}&=&\sqrt{c}\ket{W}_8+\text{e}^{i\varphi}\sqrt{1-c}\ket{\psi^{2}}_8,
\nonumber\\
\text{D7: }
\ket{\psi}_6^{\text{D7}}&=&\sqrt{c}\ket{\text{GHZ}'}_8+\text{e}^{i\varphi}\sqrt{1-c}\ket{\psi^{1}}_8,
\nonumber\\
\text{D8: }
\ket{\psi}_6^{\text{D8}}&=&\sqrt{c}\ket{\text{GHZ}'}_8+\text{e}^{i\varphi}\sqrt{1-c}\ket{\psi^{2}}_8,
\nonumber\\
\text{D9: }
\ket{\psi}_6^{\text{D9}}&=&\sqrt{c}\ket{\psi^{1}}_8+\text{e}^{i\varphi}\sqrt{1-c}\ket{\psi^{2}}_8.
\end{eqnarray}

From Table \ref{m8}, features similar to those of four- and six-qubit
cases can be found. For D1 and D9, $\Lambda_{\max}$ is obviously
dependent only on the superposition coefficients of
$\ket{\psi}_6^{\text{D1}}$ and $\ket{\psi}_6^{\text{D9}}$, and thus
$\mathcal{S}_{\ket{\text{GHZ}}_8}\sim\mathcal{S}_{\ket{\text{GHZ}'}_8}$
and
$\mathcal{S}_{\ket{\psi^{1}}_8}\sim\mathcal{S}_{\ket{\psi^{2}}_8}$.
As for D3, D5, D7,and  D8, $\Lambda_{\max}$ is also dependent only
on the superposition coefficients. However, because the common
coefficient of $\ket{\psi^{1(2)}}_8$ is $1/2$, which is smaller than that of
$\ket{\text{GHZ}}_8$ and $\ket{\text{GHZ}'}_8$,  then
$\mathcal{S}_{\ket{\text{GHZ}}_8}\sim\mathcal{S}_{\ket{\text{GHZ}'}_8}>\mathcal{S}_{\ket{\psi^{1(2)}}_8}$.
From D2, D4 and D6, $\Lambda_{\max}$ can always be obtained under
the supposition for nearest FSPs based on the TI structures of
$\ket{\text{GHZ}'}_8$ or $\ket{\psi^{1(2)}}_8$, and thus
$\mathcal{S}_{\ket{W}_8}<\mathcal{S}_{\ket{\text{GHZ}'}_8}$ and
$\mathcal{S}_{\ket{W}_8}<\mathcal{S}_{\ket{\psi^{1(2)}}_8}$.

In total one has \begin{eqnarray}
\mathcal{S}_{\ket{\text{GHZ}}_8}\sim\mathcal{S}_{\ket{\text{GHZ}'}_8}>\mathcal{S}_{\ket{\psi^{1}}_8}
\sim\mathcal{S}_{\ket{\psi^{2}}_8}>\mathcal{S}_{\ket{W}_8}.
\end{eqnarray}

\subsection{\label{sec: overlapIIF}Further discussion}
In conclusion, the hierarchy for the basic TI entangled states is a
general feature for multi-qubit states. An important observation
obtained from the studies discussed above is that this hierarchy has an important effect on
the determination of $\Lambda_{\max}$ of HTIEs.

From the above numerical evaluations, two observations can be made
\begin{enumerate}
\item $\Lambda_{\max}$ for the HTIEs, composed of
two different basic TI entangled states except of the $N$-period,
depends obviously on the superposition coefficients. That is to say,
$\Lambda_{\max}$ are just the maximal one of the superposition
coefficients times the common coefficient of the basic TI entangled
states. For example, as shown for A2 in Table \ref{m4},
$\Lambda_{\max}$ are
$\left(1-\tfrac{1}{4}\right)\times\tfrac{1}{2}$,
$\tfrac{1}{2}\times\tfrac{1}{2}$, $\tfrac{3}{4}\times\tfrac{1}{2}$
for $c=1/4, 1/2$ and $c=3/4$, respectively. More examples can be
found in C1, C9, and C11 in Table \ref{m6} and D1, D3, D5, D7, D8, and D9 in
Table \ref{m8}. This feature is consistent with the result of
Ref. \cite{chs00}.
\item The situation for the cases including $N$-period TI entangled
states is very complex, however. A general feature is that
$\Lambda_{\max}$ in this case is unrelated completely to the
superposition coefficients. However, one can find that PI is
an optimal supposition for the nearest FSPs when the HTIEs are
composed of one- and $N$-period TI entangled states, independent on
the coefficient \cite{footnote3}. For other cases our data cannot
give a clear conclusion.
\end{enumerate}

\begin{table}
\begin{tabular}{p{1.5cm} p{1.5cm} p{1.5cm} p{1.5cm} r}
\hline\hline\text{State}   &  Case 0  &  Case 1  &  Case 2  &  Case 3 \\
\hline
E1-1& {\bf 0.2407}& 0.18365 & 0.20275 & 0.18365 \\
E1-2& {\bf 0.25199}& 0.20546 & 0.21439 &0.20545 \\
E1-3& {\bf 0.24209}& 0.20251 &0.20574 &0.20251 \\ \hline\hline
\end{tabular}
\caption{\label{me1}Numerical results of the maximal overlap for the
mixed TI entangled states Eqs.\eqref{e1}. The data for case 0 are
obtained by setting $\theta_1=\theta_2=\cdots=\theta_6=-\varphi$}
\end{table}

A unique case is one of the HTIEs composed of basic $N$-period TI
entangled states, except of $\ket{W}_N$. For example, let us consider
HTIEs,
\begin{equation}\label{e1}
\ket{\psi}_6^{E1}=\sqrt{c}\ket{\psi^{1a}}_6+\text{e}^{i\varphi}\ket{\psi^{2a}}_6,
\end{equation}
of which $\Lambda_{\max}$ are listed in Table \ref{me1}. It is
obvious that $\Lambda_{\max}$ is completely independent of the TI
structures of $\ket{\psi^{1a}}_6$ and $\ket{\psi^{2a}}_6$. A similar
feature can be found also for the eight-qubit case, while our evaluation
show that this feature does not appear for D9 in Table \ref{m8},
which is the superposition of two four-period TI entangled states for
eight-qubit multipartite states. The above evaluations show that
this feature does not occur for the combination with different
periods. So we surmise that it would be unique for $N$-period basic TI
entangled states, except for $\ket{W}_N$, and one has to be careful
when dealing with this case.

Unfortunately we cannot provide the exact proofs for these
observations above, and the underlying physics is also unclear until
now. However, we should express the idea here that it would
be efficient to find the nearest
fully separable state from the TI of entangled states.

Here we discuss the numerical errors
of the random simulation of $\Lambda_{\max}$. The
numerical evaluation for $\Lambda_{\max}$ is actually an
exhaustive method: sampling randomly the permissible values for all
independent parameters as much as possible and record the maximal
value of the overlap. It is obvious that the efficiency and
accuracy of this algorithm is dependent on the sampling times
and the number of independent parameters: greater sampling times help us achieve the true result more
accurately, but the evaluation time increase. The more independent parameters we use, the slower is the convergence
of evaluation to the true result. Because the purpose of this
section is to show the effect of TI rather than to find the true
$\Lambda_{\max}$, we chose sampling times of $10^5$ in order for
a tolerable calculation time, and $\Lambda_{\max}$ is the steady
numerical result that is unchanged for $10^4$ sampling times.
We admit that the evaluation in this section is crude.
However some interesting information can still be attained from the
data listed in the previous tables. In addition, the number of
independent parameters can be reduced by some algebra, e.g.,
Eq.\eqref{m1} which makes our evaluation more reliable.

\section{summary and conclusion}
By several exact and numerical examples, we presents a comprehensive
study of the maximal overlap with fully separable state for the
multipartite entangled pure state with translational invariance. As shown in the above discussions, translational
invariance of the entangled state has an intrinsic effect on the
determination of the maximal overlap $\Lambda_{\max}$ and the
nearest fully separable state.

One contribution of this article is the introduction of the basic TI
entangled states, as shown in Sec. \ref{sec: overlapI}. The key point
for these states is that the other types of TI entangled states can
be expressed as the superposition of these basic ones. In addition
for these basic TI entangled states $\Lambda_{\max}$ and the nearest
fully separable state can be determined just by utilizing the TI.
Furthermore,  we stress that
TI or basic TI entangled states actually defines a connectedness in the
states belonging to different parties, and this connectedness is the
unique fundamental character that is distinct from one state to the other state.
So the nearest fully separable state for the entangled state should
also manifest this unique TI.

Another contribution of this paper is the demonstration of the
hierarchy for the basic TI entangled states with different periodic
structure. This hierarchy of TI has a fundamental effect on the
evaluation of $\Lambda_{\max}$ for HTIEs, as shown in Sec. \ref{sec:
overlapII} (see Sec. \ref{sec: overlapIIF} for a conclusion).

Finally, we have to admit that an exact proof for the resultant
conclusions are still absent. From our point of view it is because the
understanding of multipartite entanglement is not yet unified. As
shown in this paper, although the forms of the multipartite
entanglements are diverse, there are some fundamental features by
which one can obtain useful information, such as the
introduction of basic TI entangled states and their hierarchy. So it
is promising to understand multipartite entanglement by disclosing
these fundamental features.

\begin{acknowledgments}
One of the authors (H.T.C) thanks Prof. Dr. Xue-Xi Yi for his hospitality
during his visitto Dalian University of Technology, where the
work was finally completed. Helpful discussion with Prof. Dr. Yan-Kui Bai and Dr. Chang-Shui Yu, and critical readings by
Dr. Lin-Cheng Wang and Dr. Xiao-Li Huang are acknowledged also. This work
is supported by NSF of China, Grants No. 11005002 (Cui) and No. 11005003
(Tian).
\end{acknowledgments}

\begin{appendix}
\section{}\label{aa}
This appendix provides an exact result of $\Lambda_{\max}$ at the
general state,
\begin{equation}
\ket{W'}=c_0\ket{100}+c_1\text{e}^{i\alpha}\ket{010}+
c_2\text{e}^{i\beta}\ket{001},
\end{equation}
where $c_0,c_1,c_2\in[0,1]$, $c_1^2+c_2^2+c_3^2=1$, and $\alpha,
\beta\in[0, 2\pi]$. Then the overlap is
\begin{eqnarray}
&&\left|\inp{W'}{\phi}\right|^2\nonumber\\
&=&\left|c_0\sqrt{a_1(1-a_2)(1-a_3)}\text{e}^{i(\theta_2+\theta_3)}+\right.\nonumber\\
&&\hspace{.5em}c_1\sqrt{(1-a_1)a_2(1-a_3)}\text{e}^{i(\theta_1+\theta_3-\alpha)}+\nonumber\\
&&\hspace{.5em}\left.c_2\sqrt{(1-a_1)(1-a_2)a_3}\text{e}^{i(\theta_1+\theta_2-\beta)}\right|^2\nonumber\\
&\leq&
\left[c_0\sqrt{a_1(1-a_2)(1-a_3)}+c_1\sqrt{(1-a_1)a_2(1-a_3)}\right.\nonumber\\
&&\hspace{.5em}\left.+c_2\sqrt{(1-a_1)(1-a_2)a_3}\right]^2
\end{eqnarray}
where the equality occurs when
\begin{eqnarray}
\begin{cases}
\theta_2+\theta_3=0,\\
\theta_1+\theta_3=\alpha, \\
\theta_1+\theta_2=\beta.
\end{cases}
\end{eqnarray}
and the solutions can be easily obtained:
$\theta_1=\tfrac{\alpha+\beta}{2},
\theta_2=-\tfrac{\alpha-\beta}{2}=-\theta_3$.

Obviously one has
\begin{eqnarray}
&&\max_{\ket{\phi}}\left|\inp{W'}{\phi}\right|^2\Leftrightarrow\nonumber\\
&&\max_{a_1, a_2,a_3}
\left[c_0\sqrt{a_1(1-a_2)(1-a_3)}+c_1\sqrt{(1-a_1)a_2(1-a_3)}\right.\nonumber\\
&&\hspace{3em}\left.+c_2\sqrt{(1-a_1)(1-a_2)a_3}\right]^2.
\end{eqnarray}
Then the following discussion would focus on the expression
\begin{eqnarray}
f&=&c_0\sqrt{a_1(1-a_2)(1-a_3)}+c_1\sqrt{(1-a_1)a_2(1-a_3)}\nonumber\\
&&+c_2\sqrt{(1-a_1)(1-a_2)a_3}
\end{eqnarray}
The extremal point for $f$ corresponds to the points at which the first
derivation of $f$ with $a_1, a_2, a_3$ vanish, i.e., $\tfrac{\partial
f}{\partial a_1}=\tfrac{\partial f}{\partial a_2}=\tfrac{\partial
f}{\partial a_3}=0$. This condition is transformed into the
equations
\begin{eqnarray}\label{a}
\begin{cases}
c_1\sqrt{\frac{a_2}{1-a_2}}+c_2\sqrt{\frac{a_3}{1-a_3}}=c_0\sqrt{\frac{1-a_1}{a_1}},\\
c_0\sqrt{\frac{a_1}{1-a_1}}+c_2\sqrt{\frac{a_3}{1-a_3}}=c_1\sqrt{\frac{1-a_2}{a_2}},\\
c_0\sqrt{\frac{a_1}{1-a_1}}+c_1\sqrt{\frac{a_2}{1-a_2}}=c_2\sqrt{\frac{1-a_3}{a_3}}.
\end{cases}
\end{eqnarray}
Introducing the media variables
$x=\sqrt{\tfrac{a_1}{1-a_1}\tfrac{a_2}{1-a_2}},
y=\sqrt{\tfrac{a_2}{1-a_2}\tfrac{a_3}{1-a_3}},
z=\sqrt{\tfrac{a_1}{1-a_1}\tfrac{a_3}{1-a_3}}$, the
 above equation can be rewritten as
\begin{eqnarray}
\begin{cases}
c_1 x+ c_2 z=c_0,\\
c_0 x+c_2 y=c_1,\\
c_0 z+c_2 y=c_2.
\end{cases}
\end{eqnarray}
where the solutions are $x=\tfrac{1-2c_2^2}{2c_0c_1},
y=\tfrac{1-2c_0^2}{2c_1c_3}, z=\tfrac{1-2c_1^2}{2c_0c_2}$.
Furthermore by evaluating the ratios $\frac{x}{y},
\frac{x}{z},\frac{y}{z}$ and with the constraint
$c_0^2+c_1^2+c_2^2=1$, one can completely determine the values of $c_0,
c_1, c_2$.

The next step is to judge whether the point corresponds to the maximally
extremal point, where $f$ has to satisfy the relation
$\tfrac{\partial^2f}{\partial a_{\alpha}\partial a_{\beta}}\leq0
(\alpha,\beta=1,2,3)$. With Eqs. \eqref{a}, one can check out easily
$\tfrac{\partial^2f}{\partial a_{\alpha}\partial a_{\beta}}\leq0
(\alpha\neq\beta)$. Thus one has
\begin{eqnarray}
\begin{cases}
\frac{\partial^2f}{\partial
a_1^2}=\frac{\sqrt{(1-a_2)(1-a_3)}}{4(1-a_1)^{3/2}}c_0
\sqrt{\frac{1-a_1}{a_1}}(2-\frac{1}{a_1}),\\
\frac{\partial^2f}{\partial
a_2^2}=\frac{\sqrt{(1-a_1)(1-a_3)}}{4(1-a_2)^{3/2}}c_1
\sqrt{\frac{1-a_2}{a_2}}(2-\frac{1}{a_2}),\\
\frac{\partial^2f}{\partial
a_3^2}=\frac{\sqrt{(1-a_1)(1-a_2)}}{4(1-a_3)^{3/2}}c_2
\sqrt{\frac{1-a_3}{a_3}}(2-\frac{1}{a_3}).
\end{cases}
\end{eqnarray}
where Eq.\eqref{a} is used. It is obvious that only if
$a_\alpha<\frac{1}{2}$, then $\frac{\partial^2f}{\partial
a_{\alpha}^2}\leq0$. With this condition one can easily find the
maximally extremal values of $f$.

Finally it should be pointed that the maximal values are in general
not completely consistent with the extremal values. Thus the
boundary points $a_{\alpha}=0,1$ must be checked independently.

\end{appendix}


\begin{thebibliography}{99}
\bibitem{footnote0}The meaning of the next-nearest-neighbor site in this article
is defined by the sequence of all party labels with geometry shown
in Fig. \ref{1}.
\bibitem{footnote2} As for $\ket{\overline{W}}_4=\tfrac{1}{2}(\ket{0111}+\ket{1011}+
\ket{1101}+\ket{1110})$, it is obvious that
$\ket{\overline{W}}_4=\otimes_{n=1}^4\sigma_{n}^x\ket{W}$. Thus
$\ket{\overline{W}}_4$ is equivalent to $\ket{W}$, and in addtion
$\ket{\overline{W}}_4$ has also the same TI to $\ket{W}$ because
$\sigma_{n}^x$ operates every party at the same time. Thus it is not
necessary to discuss this case solely. This principle is also
applied in the following discussion; If two entangled states could
be transformed into each other by the local unitary operation, which
is the tensor product of the same one unitary singl-party operator,
e.g. $\otimes_{n=1}^4\sigma_{n}^x$ for the four-qubit case, only one of
them will be discussed in this paper. It should be pointed out that
$\ket{\text{GHZ}'}_N$ thus has a different TI from
$\ket{\text{GHZ}}_N$ because the operation
$\sigma^x_2\otimes\sigma^x_4\otimes\cdots\otimes\sigma^x_{2n}\otimes\cdots$
is only imposed on the states in even site.
\bibitem{footnote3}This point becomes clearer for the case including
$\ket{W}_N$ because $\ket{W}_N$ has the same common coefficient
$1/\sqrt{N}$ to that of basic $N$-period TI entangled states, which
excludes the effect of the superposition coefficients.
\end{thebibliography}
\end{document}